\documentclass[twocolumn]{aastex7}
\usepackage{amsmath}
\usepackage{longtable}

\begin{document}
\title{Search for Slow Bars in Two Barred Galaxies with Nuclear Structures: NGC 6951 and NGC 7716}

\author[0000-0003-2779-6793]{Yun Hee Lee}
\affil{Department of Astronomy and Atmospheric Sciences, Kyungpook National University, Daegu 41566, Republic of Korea}
\affil{Korea Astronomy and Space Science Institute (KASI), 776 Daedeokdae-ro, Yuseong-gu, Daejeon 34055, Republic of Korea}
\email[show]{yhinastro@gmail.com}  

\author[0000-0003-3428-7612]{Ho Seong Hwang}
\affil{Astronomy Program, Department of Physics and Astronomy, Seoul National University, 1 Gwanak-ro, Gwanak-gu, Seoul 08826, Republic of Korea}
\affil{SNU Astronomy Research Center, Seoul National University, 1 Gwanak-ro, Gwanak-gu, Seoul 08826, Republic of Korea}
\affil{Korea Astronomy and Space Science Institute (KASI), 776 Daedeokdae-ro, Yuseong-gu, Daejeon 34055, Republic of Korea}
\affiliation{Australian Astronomical Optics - Macquarie University, 105 Delhi Road, North Ryde, NSW 2113, Australia}
\email{hhwang@astro.snu.ac.kr}  

\author[0000-0001-9922-583X]{Virginia Cuomo}
\affil{Departamento de Astronomía, Universidad de La Serena, Av. Raúl Bitrán 1305, La Serena, Chile}
\email{virginia.cuomo@userena.cl}

\author[0000-0003-1544-8556]{Myeong-Gu Park}
\affil{Department of Astronomy and Atmospheric Sciences, Kyungpook National University, Daegu 41566, Republic of Korea}
\email{mgp@knu.ac.kr}

\author[0000-0002-5857-5136]{Taehyun Kim}
\affil{Department of Astronomy and Atmospheric Sciences, Kyungpook National University, Daegu 41566, Republic of Korea}
\email{tina1739@gmail.com}

\author[0000-0002-2013-1273]{Narae Hwang}
\affil{Korea Astronomy and Space Science Institute (KASI), 776 Daedeokdae-ro, Yuseong-gu, Daejeon 34055, Republic of Korea}
\email{nhwang@kasi.re.kr}

\author[0000-0002-9822-5608]{Hong Bae Ann}
\affiliation{Department of Earth Science Education, Pusan National University, Busan 46241, Republic of Korea}
\email{hbann@pusan.ac.kr}

\author[0000-0003-4625-229X]{Woong-Tae Kim}
\affil{Astronomy Program, Department of Physics and Astronomy, Seoul National University, 1 Gwanak-ro, Gwanak-gu, Seoul 08826, Republic of Korea}
\affil{SNU Astronomy Research Center, Seoul National University, 1 Gwanak-ro, Gwanak-gu, Seoul 08826, Republic of Korea}
\email{wkimism@gmail.com}

\author[0000-0001-9263-3275]{Hyun-Jeong Kim}
\affil{Korea Astronomy and Space Science Institute (KASI), 776 Daedeokdae-ro, Yuseong-gu, Daejeon 34055, Republic of Korea}
\email{hjkim@kasi.re.kr}

\author[0000-0002-0070-3246]{Ji Yeon Seok}
\affil{Korea Astronomy and Space Science Institute (KASI), 776 Daedeokdae-ro, Yuseong-gu, Daejeon 34055, Republic of Korea}
\email{jyseok@kasi.re.kr}

\author[0000-0003-3301-759X]{Jeong Hwan Lee}
\affil{Astronomy Program, Department of Physics and Astronomy, Seoul National University, 1 Gwanak-ro, Gwanak-gu, Seoul 08826, Republic of Korea}
\affil{SNU Astronomy Research Center, Seoul National University, 1 Gwanak-ro, Gwanak-gu, Seoul 08826, Republic of Korea}
\email{joungh93@gmail.com}

\author[0009-0005-5145-5165]{Yeon-Ho Choi}
\affil{Korea Astronomy and Space Science Institute (KASI), 776 Daedeokdae-ro, Yuseong-gu, Daejeon 34055, Republic of Korea}
\affil{Korea National University of Science and Technology, Daejeon 34113, Republic of Korea}
\email{yhc@kasi.re.kr}

\begin{abstract}
We investigate two barred galaxies with nuclear structures, NGC 6951 and NGC 7716, to examine whether they host slow bars. Using Gemini/GMOS long-slit spectroscopy, we calculate the bar pattern speed with the Tremaine-Weinberg method and detect kinematically decoupled nuclear disks in both galaxies. We also measure the bar length and strength using Pan-STARRs images and identify a nuclear ring in NGC 6951 and a nuclear bar in NGC 7716 from HST/PC images. Our results indicate that NGC 6951 hosts a slow, long, and strong bar, which likely evolved through interactions with the dark matter halo and contributed to the formation of both the nuclear disk and ring. We also find hints of a rapidly rotating oval structure within the primary bar, although it is not clearly seen in the imaging data. In contrast, the primary bar in NGC 7716 is too weak to be classified as a barred galaxy, while its nuclear disk and nuclear bar are unusually large, possibly due to tidal interactions or the weakness of the primary bar. These findings suggest that slow bars may be more observed in galaxies with nuclear structures and highlight the often underappreciated role of galaxy interactions in bar evolution.
\end{abstract} 

%Theoretically, nuclear structures such as nuclear rings and nuclear bars are expected to be associated with the inner Lindblad \textcolor{orange}{r}esonance (ILR), which, in turn, is linked to the presence of slow bars.
%% to include these concepts in their preprints.$$


\section{Introduction} \label{chap1} %===================================

Stellar bars are a common feature in galaxies, observed in approximately 60\% of disk galaxies in the local universe \citep{1991deVaucouleurs, 2008Sheth, 2015Ann, 2015Buta, 2019Lee}. Their widespread occurrence can be attributed to diverse formation pathways: bars can spontaneously emerge in isolated galaxies with cold disks due to stellar disk instabilities \citep{1964Toomre, 1973Ostriker}, or can be induced by tidal interactions \citep{1998Miwa, 2014Lokas, 2018Lokas, 2017Martinez_Valpuesta, 2019Yoon}. 
%However, this complexity makes it challenging to fully understand their formation and evolutionary processes, both in simulations and observations.}

In isolated systems, interactions between bars and dark matter halos are considered the primary driver of bar evolution. Most simulation studies suggest that, as a result of dynamical friction with bulges or dark matter halos, bars gradually slow down over time, accompanied by an increase in both their length and strength \citep{1980Sellwood, 2000Debattista, 2002Athanassoula, 2003Athanassoula, 2014Athanassoula, 2023Jang, 2024Jang}. These findings are further supported by cosmological hydrodynamical simulations based on $\Lambda$CDM, including the Evolution and Assembly of GaLaxies and their Environments (EAGLE; \citealp{2015Schaye}) and the IllustrisTNG \citep{2019Nelson}, which show that strong bars slow down rapidly and predominantly exhibit slow rotation at z $\sim$ 0 \citep{2017Algorry, 2021Roshan}. On the other hand, the Auriga simulations \citep{2017Grand}, which involve more baryon-dominated galaxies, show that most bars remain fast down to $z=0$ \citep{2021Fragkoudi}. This discrepancy also suggests that the dark matter halo may play a crucial role in bar evolution. 

Therefore, in observations, measuring three parameters - bar pattern speed, bar length, and bar strength - is critical for investigating bar evolution: bar pattern speed is determined through spectroscopic observations \citep[hereafter TW method]{1984Tremaine}, while bar length and strength are derived from photometric data \citep{1990Ohta, 2004Laurikainen, 2004Jogee, 2019Lee, 2020Lee}. Although the TW method was first introduced in the 1980s to directly measure bar pattern speed using long-slit spectroscopic observations covering multiple positions for a single galaxy \citep{1984Tremaine, 1987Kent, 1995Merrifield}, it has recently become applicable to large samples of galaxies thanks to the advent of Integral Field Spectroscopy (IFS), which allows multiple pseudo-long-slits to be generated from a single observation \citep{2015Aguerri, 2019Guo, 2019Cuomo, 2020Garma-Oehmichen, 2022Garma-Oehmichen, 2023Geron}. 

However, observations have not confirmed the bar evolution driven by interactions with dark matter halos as predicted by simulations. Barred galaxies in the local universe predominantly host fast bars \citep{2015Aguerri, 2019Guo, 2019Cuomo, 2020Garma-Oehmichen, 2022Garma-Oehmichen}. Moreover, bar evolution in terms of changes in bar pattern speed, length, and strength has rarely been detected \citep{2012Perez, 2021Kim, 2022Lee}. These observational results have raised questions about the amount of dark matter in the universe \citep{2021Roshan, 2023Romeo, 2023Nagesh}.

%Bar evolution is likely a more complex and multiparametric \textcolor{orange}{process} influenced by \textcolor{orange}{various} factors\textcolor{orange}{, such as} the amount of gas \citep{2014Athanassoula, 2019Seo, 2023Beane}, \textcolor{orange}{as well as} the shape and spin of the halo \citep{2014Athanassoula, 2014Long, 2018Collier, 2024Jang}, among others. 

On the other hand, more recently, \citet{2023Geron} reported a large sample of slow bars, identifying 62\% of 225 barred galaxies, based on data from the Mapping Nearby Galaxies at APO (MaNGA; \citealp{2015Bundy}). Their sample is characterized by a significant number of weak bars, which tend to appear as slow bars. Similarly, using the Multi-Unit Spectroscopic Explorer (MUSE; \citealp{2010Bacon}), several slow bars are detected in dwarf barred galaxies, which are short and weak \citep{2022Buttitta, 2022Cuomo, 2024Cuomo}. These weak and slow bars may be better explained by galaxy interactions rather than by interactions with dark matter halos \citep{2022Cuomo, 2024Cuomo, 2023Geron}. 

In interacting systems, bars are known to be born slow and remain slow throughout their evolution \citep{1998Miwa, 2016Martinez-Valpuesta, 2017Martinez_Valpuesta, 2016Lokas, 2018Lokas}. Recent studies suggest that this is likely because tidal interactions are the only way for galaxies with stable disks to form bars \citep{2025Zheng}, rather than due to bar slowdown caused by galaxy interactions \citep{2024Jimnez-Arranz}. Such bars inherently rotate more slowly than those formed in cold disks.

%Therefore, to piece together the puzzle of bar evolution and achieve a consistent understanding between simulations and observations, a much greater number of observational studies on bar properties, with a variety of approaches, \textcolor{orange}{are} needed.
%\textcolor{orange}{These galaxies are likely embedded in massive, centrally concentrated dark matter halos \citep{2014Adams, 2019Relatores}.}

%This study aims to contribute to ongoing efforts to detect typical slow bars that evolve through interactions with dark matter halos \citep{2000Debattista, 2002Athanassoula, 2014Athanassoula, 2023Jang}. 
As a result, there is still a lack of slow bar samples that have evolved through interactions with dark matter halos. In this study, we focus on measuring bar pattern speeds in barred galaxies with complex nuclear structures, such as nuclear rings or nuclear bars. Theoretically, slow bars are related to the inner Lindblad resonance (ILR; \citealp{1984Schwarz, 1985Combes, 1994Byrd}). As bar pattern speed decreases, additional resonances appear sequentially, from the outer Lindblad resonance (OLR) to the corotation radius (CR), then the ultraharmonic resonance (UHR), and finally the ILR. Nuclear features such as rings or bars are thought to require the presence of the ILR \citep{1985Combes, 1986Buta, 1994Byrd, 1996Combes}. Therefore, barred galaxies with nuclear rings or bars may serve as excellent candidates for detecting slow bars that have evolved through interactions with dark matter halos.

%This work contributes to the ongoing efforts to understand the evolution of slow bars in galaxies, particularly through interactions with dark matter halos \citep{2000Debattista, 2002Athanassoula, 2014Athanassoula, 2023Jang}.

 Here, we examine two barred galaxies, NGC 6951 with a nuclear ring and NGC 7716 with a nuclear bar. These galaxies were chosen from the Atlas of Images of NUclear Rings (AINUR; \citealp{2010Comeron}), based on the geometric conditions necessary to measure bar pattern speeds using the TW method \citep{1984Tremaine, 2019Guo}. To determine the three key parameters of the bars, long-slit spectroscopic observations were conducted using Gemini-North/GMOS, and photometric images were obtained from the Hubble Space Telescope (HST) Wide Field Planetary Camera (WFPC) and the Pan-STARRs DR1 (PS1) archives. Throughout this paper, we use a Hubble constant of $H_0=68.4~\rm kms^{-1}Mpc^{-1}$ based on the Plank 2015 results \citep{2016Plank}.

This paper is organized as follows. Section \ref{chap2} introduces the TW method for measuring bar pattern speed and describes the long-slit observations. Section \ref{chap3} represents the nuclear features of the target galaxies revealed by high-resolution photometry and spectroscopy. In Section \ref{chap4}, we analyze bar properties, including length ($R_{\rm bar}$), strength ($S_{\rm bar}$), pattern speed ($\Omega_{\rm bar}$), corotation radius ($R_{\rm CR}$), and rotation rate ($\mathcal{R}$). Finally, Sections \ref{chap5} and \ref{chap6} provide the discussion and conclusions, respectively.

%\citet{2017Algorry} concerned the numerical resolution.
\section{Method and Observations} \label{chap2} %===================================
\subsection{Tremaine-Weinberg method} \label{chap2.1} 

The pattern speed of bars, $\Omega_{\rm bar}$, is a fundamental quantity for understanding the dynamics and evolution of barred disk galaxies \citep{1987Binney, 2000Debattista}. As of now, various methods have been proposed to measure $\Omega_{\rm bar}$, utilizing photometric, spectroscopic, or modeling techniques, each of which is based on distinct characteristics or assumptions. For instance, some approaches explore the relation between resonance locations and galaxy rings \citep{1982Athanassoula, 2004Munoz-Tunon, 2012Perez}, while others examine the connection between periodic orbits and gas flow \citep{1982vanAlbada, 1992Athanassoula}. Additionally, variations in galaxy properties near the corotation radius, $R_{\rm CR}$, have also been investigated, including changes in stellar population \citep{1997Puerari, 2009Buta, 2013Scarano}, spiral morphology \citep{1993Canzian, 2024Marchuk}, and streaming motion \citep{2011Font, 2014Font}. However, despite these advancements, consistent results across different methods remains elusive, making the reliable measurement of $\Omega_{\rm bar}$ difficult \citep{2024Kostiuk, 2024Marchuk}.  

\begin{figure*}
\centering
\includegraphics[width=0.75 \linewidth, trim = {180 10 250 0}, clip]{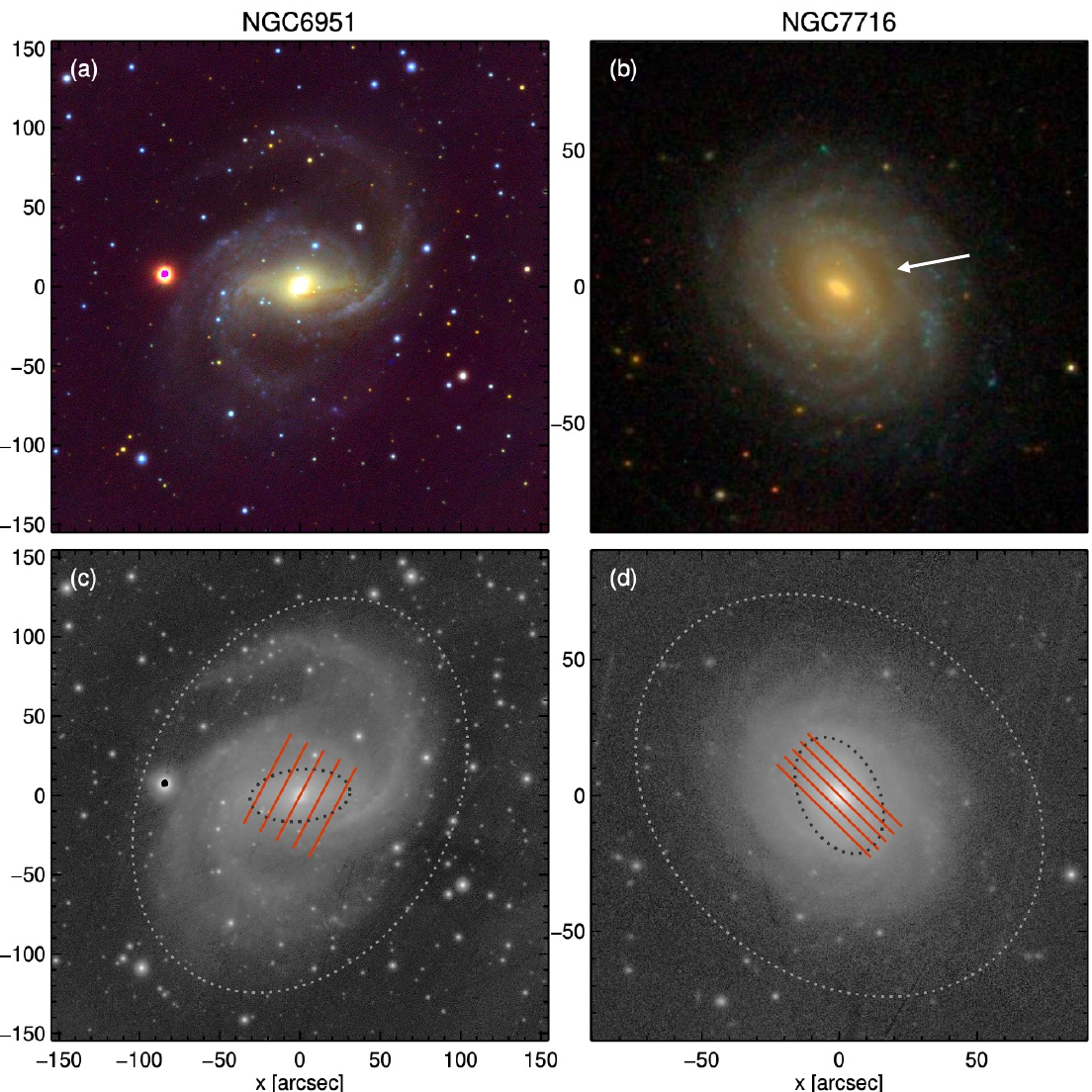}
\caption{Color images of the target galaxies are shown in the top panels: NGC 6951 (left) from PS1 and NGC 7716 (right) from SDSS. The white arrow in NGC 7716 (b) indicates the bar identified in some previous studies \citep{2006Knapen, 2010Comeron}, but the initial classification describes a weak bar surrounded by the inner ring \citep{1980deVaucouleurs}. The bottom panels display the shapes of the disks and bars, constructed using the PA and ellipticity from BarIstA, with large and small gray dotted ellipses overlaid on the PS1 $i-$band images. The red lines indicate the positions of five long-slits, aligned with the major axis of the disks, and their lengths correspond to regions where the S/N exceeds 3 per pixel in the observations.}
\label{Fig1}
\end{figure*}

Among these available methods, the TW method stands out as the only model-independent and direct approach for measuring $\Omega_{\rm bar}$. It has been applied more extensively than any other method across a large number of galaxy samples \citep{2015Aguerri, 2019Guo, 2019Cuomo, 2022Cuomo, 2024Cuomo, 2020Garma-Oehmichen, 2022Garma-Oehmichen, 2023Geron}. This method assumes a well-defined pattern speed and requires that tracers, such as stars, satisfy the continuity equation \citep{1984Tremaine}. $\Omega_{\rm bar}$ is derived from the line-of-sight velocity distribution (LOSVDs) through multiple long-slit spectroscopic observations, as shown in the bottom panels of Figure \ref{Fig1}, where the slit positions are marked in red. The slits are positioned along the major axis of the disk, spanning across the bar. For each slit, when the luminosity-weighted velocity and position are integrated, the contributions from the disk are canceled out, leaving only the asymmetries introduced by the bar, which provide information about the rotation of the bars \citep{2019Zou}. Further details will be discussed in Section \ref{chap4.2}.

\subsection{Target Selection and Orientation Parameters} \label{chap2.2} %===================================
We selected our target galaxies from the AINUR catalog \citep{2010Comeron}, which includes 107 nearby galaxies identified as hosting nuclear rings. These galaxies, located within 80 Mpc, are well-suited for detailed studies of nuclear structures using high-resolution HST images. From the 78 barred galaxies, we chose two targets, NGC 6951 and NGC 7716, as shown in Figure \ref{Fig1}, which satisfy the following criteria and have observational visibility. The selection criteria required galaxies to have a moderate inclination of $20^\circ < i < 60^\circ$ and an offset between the position angles ($\rm PAs$) of the bar and the disk, with $10^\circ < |\rm PA_{disk} - PA_{bar}| < 80^\circ$ \citep{2015Aguerri, 2019Zou, 2019Guo, 2019Cuomo, 2020Garma-Oehmichen}. These are the necessary geometric conditions for applying the TW method, avoiding configurations where bar contributions appear symmetric with respect to the disk \citep{2019Zou}.

%Nuclear rings were identified through structure maps \citep{2002Pogge}, color maps, and continuum-subtracted H$\alpha$ and Pa$\alpha$ images \citep{2004Knapen, 2006Knapen, 2008Comeron}, along with UV images from the HST archive \citep{2010Comeron}. %\textcolor{orange}{On the other hand, these galaxies have been excluded from IFS surveys due to their larger angular sizes, which exceed the field of view (FOV) of such surveys.}

Therefore, accurate measurements of the galaxy disk orientation parameters ($\rm PA_{disk}$ and ellipticity, $\epsilon_{\rm disk}$) as well as $\rm PA_{bar}$ are essential for selecting targets. In particular, errors in $\rm PA_{disk}$ are the dominant source of systematic uncertainty in the application of the TW method \citep {2003Debattista, 2019Zou, 2020Garma-Oehmichen}, and more generally, inaccurate orientation parameters significantly affect bar property measurements, including bar length, strength, and pattern speed \citep{2010Comeron, 2020Lee, 2020Garma-Oehmichen, 2023Geron}. This issue becomes especially critical when dealing with large samples, where inaccurate orientation parameters can lead to misclassification of bars or introduce major uncertainties in bar property measurements \citep{2020Lee}. However, by visually inspecting the variation of isophotes for each galaxy, potential misjudgments can be significantly reduced.

Orientation parameters are typically derived from isophotes in the sky plane using ellipse fitting \citep{2002Jogee, 2002Laine, 2015Aguerri, 2019Cuomo, 2019Guo}. In previous studies, the reference isophote has been chosen in different ways, such as using the 25 mag/arcsec$^2$ surface brightness contour \citep{2002Jogee, 2002Laine, 2019Lee} or averaging over the disk region where the PA and $\epsilon$ appear roughly constant \citep{2019Cuomo, 2019aCuomo}. In this study, we determined the orientation parameters of the galaxy disks and $\rm PA_{bar}$ by performing an IDL-based ellipse fitting on sky-subtracted PS1 i-band images with bright stars masked, using the Bar Image astro-Arithmometer \citep[BarIstA\footnote{This is available from \url{https://github.com/yhinastro/BarIstA}};][]{2019Lee, 2020Lee}. The masking was performed automatically using a built-in BarIstA routine, which identifies and removes foreground stars or nearby galaxies by analyzing the spatial variation in brightness across the image. The disk orientation parameters were measured as the mean PA and $\epsilon$ within the radial range where both profiles remain approximately constant. The inclination was derived from the mean $\epsilon$ using $i = \arccos(1 - \epsilon)$, and the $\rm PA_{bar}$ was measured at the radius of maximum $\epsilon$. The isophotes used for the disk and the bar are shown as large and small gray dotted ellipses, respectively, in the bottom panels of Figure \ref{Fig1}. These measurements, along with the basic properties of the target galaxies, are listed in Table \ref{Table1}. The standard deviations of the $\rm PA_{disk}$ within this selected region are $1.7^{\circ}$ for NGC 6951 and $1.3^{\circ}$ for NGC 7716, and are adopted as the uncertainties in the PA measurements.

%%---- Table1 ---------------
\begin{deluxetable*}{cccccccccc}
 \tablecaption{Properties of target galaxies \label{Table1}}
 \tablehead{
 \colhead{Galaxy} &
 \colhead{RA} &
 \colhead{Dec.} &
 \colhead{$v_{\rm sys}$} &
 \colhead{Distance} &
 \colhead{Morphology} &
 \colhead{$\rm PA_{disk}$} &
 \colhead{$\rm PA_{bar}$} &
 \colhead{Incl.}\\
 \colhead{ } &
 \colhead{(hh mm ss)} &
 \colhead{($^{\circ}$ $'$ $''$)} &
 \colhead{($\rm km~s^{-1}$)} &
 \colhead{($\rm Mpc$)} &
 \colhead{ } &
 \colhead{($^{\circ}$)} &
 \colhead{($^{\circ}$)} &
 \colhead{($^{\circ}$)}\\
  \colhead{(1) } &
 \colhead{(2)} &
 \colhead{(3)} &
 \colhead{(4)} &
 \colhead{(5)} &
 \colhead{(6) } &
 \colhead{(7)} &
 \colhead{(8)} &
 \colhead{(9)}\\
}
\startdata
NGC 6951 & 20 37 14 & +66 06 20 & 1411$\pm$2 & 20.6 & SAB(rs)bc & 152.5$\pm$1.7 & 96.0 & 43.0 \\
NGC 7716 & 23 36 31 & +00 17 50 & 2556$\pm$2 & 37.4 & SAB(r)b   & 45.0$\pm$1.3  & 24.8 & 38.2\\ 
\enddata
\tablecomments{\footnotesize Galaxy name (Column 1), position on the sky in J2000.0 coordinates (Columns 2 and 3), heliocentric systemic velocity (Column 4) and distance (Column 5) measured in this study, morphological type from NED (Column 6), position angles of the disk (Column 7) and the bar (Column 8), and inclination (Column 9) measured in this study. PA is measured from North to East.} 
\end{deluxetable*}
%\vspace{-7mm} % 간격을 5mm 줄임

 %----------------------------

\subsection{Gemini/GMOS Long-slit Observations and Data Reduction} \label{chap2.3} %===================================

NGC 6951 and NGC 7716 were observed using long-slit spectroscopy with the GMOS instrument on Gemini-North (Program ID: GN-2023B-Q-303; PI: Y.H. Lee). Since the field of view of GMOS in IFU mode is limited to $5'' \times 7''$, we adopted long-slit mode to achieve spatially resolved spectroscopy over a wider area. For each galaxy, five slits were positioned along the major axis of the disk for each galaxy, as shown in red in the bottom panels of Figure \ref{Fig1}: one at the center and four symmetrically placed across the bar. For NGC 6951, the slits were aligned at a PA of 152.5$^\circ$ along the disk major axis, offset by $11.\!\arcsec45$ across the bar major axis. For NGC 7716, the slits were aligned with a PA of 225.0$^\circ$ with offsets of $4.\!\arcsec0$. 
%\textcolor{orange}{Each GMOS slit spans a length of $5.5'$.}
%The distances to these galaxies are 24.4 Mpc for NGC 6951 and 36.6 Mpc for NGC 7716 \citep{2010Comeron}. 

%\textcolor{orange}{corresponding to an FWHM of} $\sim3$~\text{\AA} at 8600~\text{\AA}.

The TW method requires a wavelength window free of strong emission lines in order to trace the stellar kinematic signal reliably \citep{2024Cuomo}. Absorption features such as the Mg I triplet (5167, 5173, and 5184 \text{\AA}) have been commonly used for this purpose \citep{2019aCuomo, 2022Buttitta, 2024Cuomo}. Recently, \citet{2024Cuomo} demonstrated that the Ca II triplet (8498, 8542, and 8662 \text{\AA}) yields consistent results with those obtained using the Mg I triplet, showing that the TW method provides robust measurements regardless of the choice of absorption lines. Considering the grating efficiency and spectral resolution of GMOS, we selected the Ca II region and used the R831 grating centered at 8600~\text{\AA}, covering a wavelength range of 7411$-$9892~\text{\AA}. A slit width of $0.\!\arcsec75$ was employed, yielding a spectral resolution of $\sim2900$, corresponding to an FWHM of $\sim$~3~\text{\AA} at 8600~\text{\AA}. The long-slit spans a length of $5.\!'5$ with a spatial sampling of 0.1614 arcsec$~\!$pixel$^{-1}$. Given the observational conditions\footnote{The observations were conducted under band 3 observing conditions: 85th percentile image quality, 80th percentile sky background, and 70th percentile cloud coverage.}, we achieved S/N $\sim3$ in regions with $\mu \sim 21~\!\mathrm{mag/arcsec^2}$, indicated by the length of the red lines in the bottom panels in Figure \ref{Fig1}. This slit length is sufficiently long to obtain a reliable $\Omega_{\rm bar}$, which will be discussed in Section \ref{chap4.2}. The exposure times per slit were 5300 and 5425 seconds for NGC 6951 and NGC 7716, respectively, totaling approximately 7.5 hours for each galaxy.

%\textcolor{orange}{We aimed the} exposure time to achieve a signal-to-noise ratio (S/N) of about 3 per pixel in regions with a disk surface brightness ($\mu$) of 22$~\!\rm mag/arcsec^2$. 

We reduced the data using the Data Reduction for Astronomy from Gemini Observatory North and South \citep[\textit{Dragons;}][]{2019Labrie}, incorporating bias subtraction, flat-field correction, and wavelength calibration using arc exposures. Flux calibration was achieved using standard star observations. The exposures were dithered along the spectral directions, and the final 2D exposures were combined to remove artifacts and bad pixels. 

\begin{figure*}
\centering
\includegraphics[width= 0.8 \linewidth, trim = {0 170 15 10}, clip]{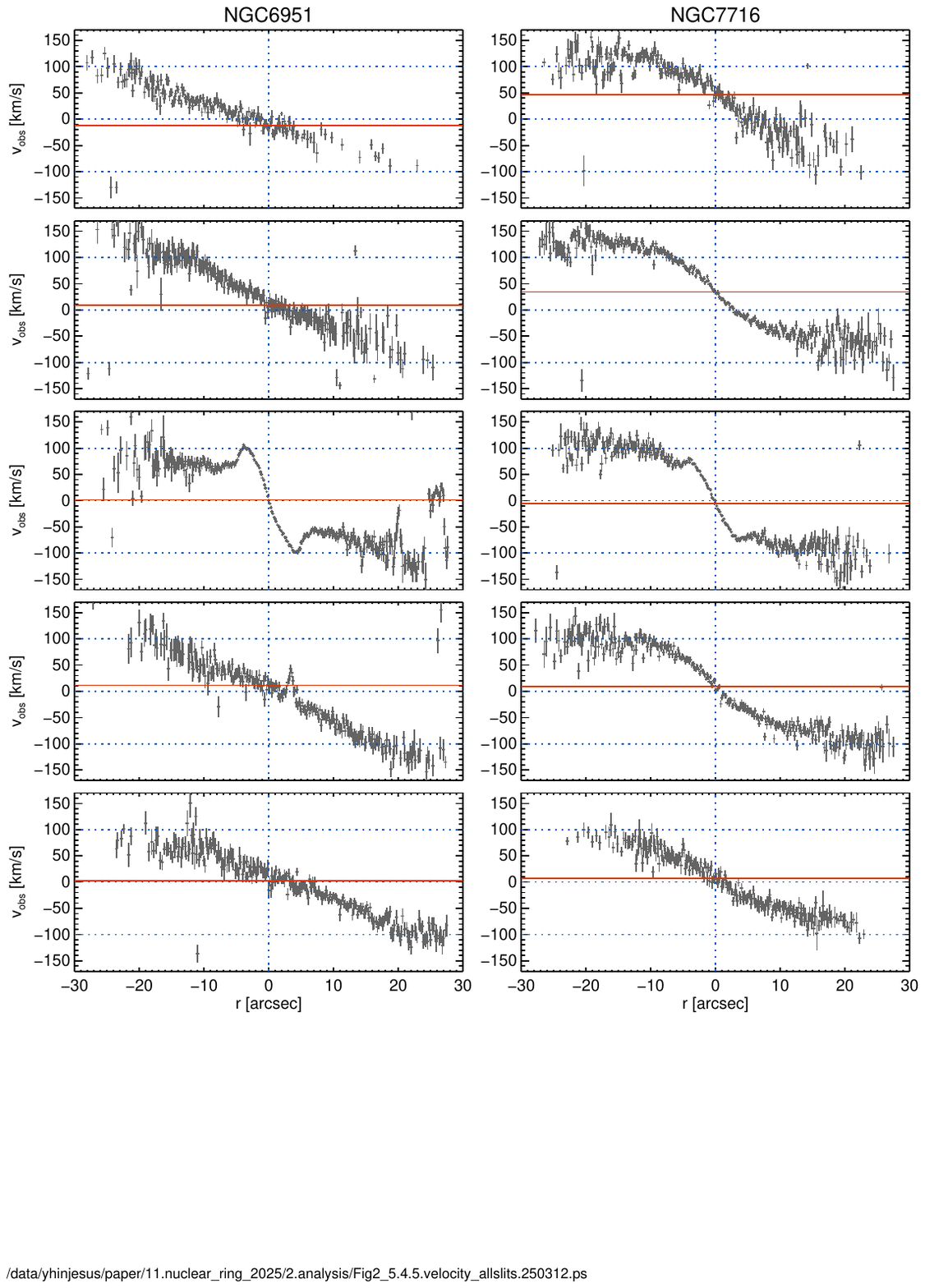}
\caption{Profiles of observed velocity from five long-slit spectra for NGC 6951 (left) and NGC 7716 (right). From top to bottom, the panels correspond to the slits positioned from left to right in the bottom panels of Figure \ref{Fig1}, with the third panels representing the slits along the major axes of each galaxy. Velocities are derived by fitting Gaussian profiles to the Ca II triplet lines in Voronoi-binned spectra, ensuring an S/N $>$ 3. Uncertainties are estimated using chi-square minimization. Blue dotted and red solid line are included as visual guides for comparison, with the red solid lines representing the mean velocities within the central 5 pixels of the third slits for each galaxy along the major axes.}
\label{FigA}
\end{figure*}

Figure \ref{FigA} shows the velocity profiles derived from the five long-slit spectra for NGC 6951 (left panel) and NGC 7716 (right panel). From top to bottom, the panels correspond to the slits arranged from left to right in the bottom panels of Figure \ref{Fig1}. The spectra have been sky-subtracted and Voronoi-binned along the spatial direction to achieve an S/N $>$ 3 \citep{2003Cappellari}. Velocities were measured by fitting Gaussian profiles to the Ca II triplet lines, with uncertainties estimated using chi-square minimization. We will discuss these results in Section \ref{chap4.2.2}.

%We examine the velocity distributions from the five long-slit spectra to determine whether the velocity integrals (Figures \ref{Fig8}(b) and (e)) result from specific noise source or not. 

\section{Nuclear Structures} \label{chap3} %===================================
\subsection{Nuclear Features from Photometry} \label{chap3.1} 

\begin{figure*}
\centering
\includegraphics[width= \linewidth, trim = {5 40 0 30}, clip]{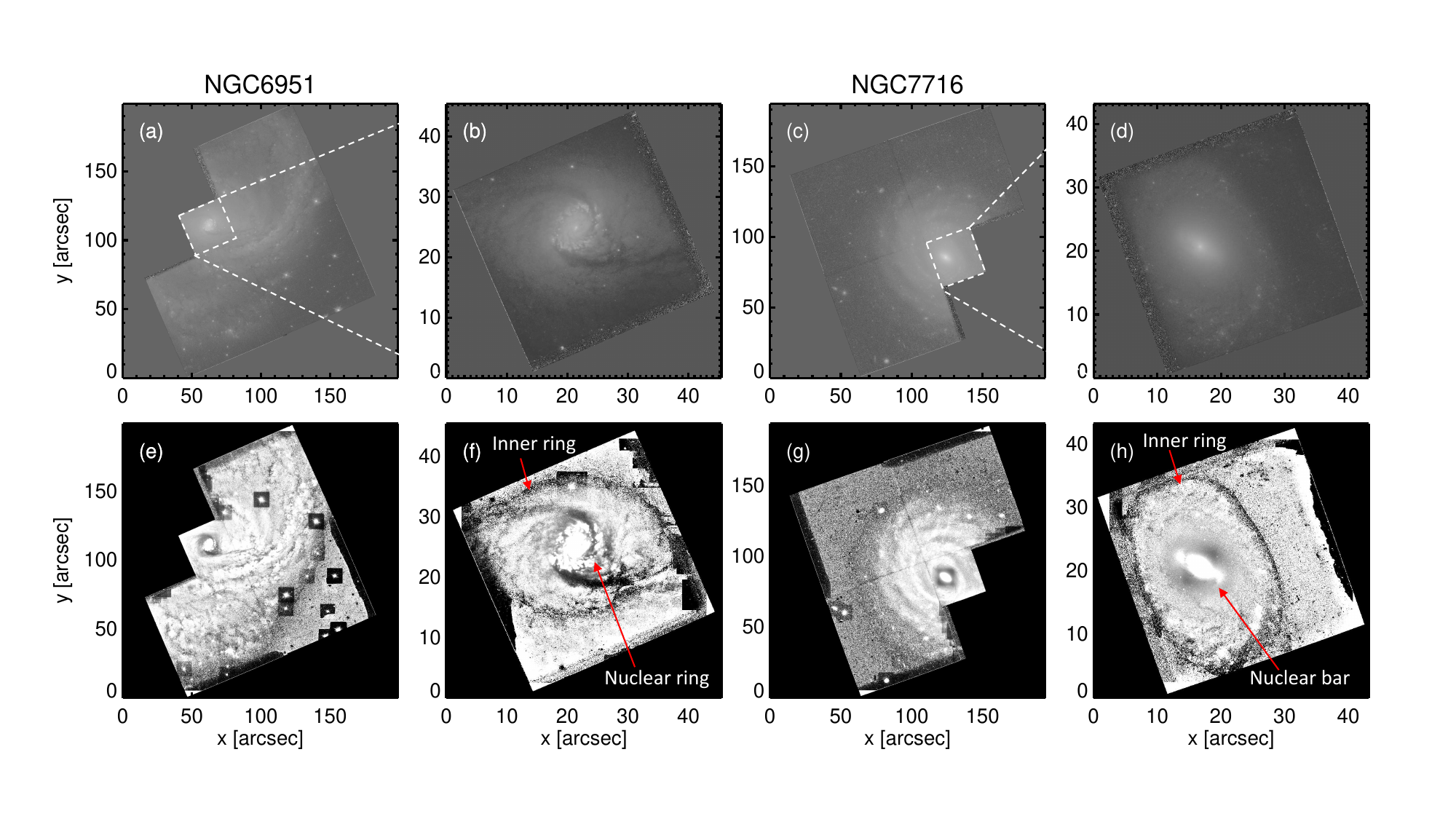}
\caption{Nuclear structures of NGC 6951 (left two columns) and NGC 7716 (right two columns) observed in F606W images from HST WF (left) and PC (right) cameras. The bottom panels highlight the nuclear features using unsharp mask processing, corresponding to each of the top four panels. The images reveal a nuclear ring and bar, as well as inner rings that surround the bar structures.}
\label{Fig2}
\end{figure*}

\begin{figure}
\centering
\includegraphics[width= 0.95\linewidth, trim = {10 160 310 260}, clip]{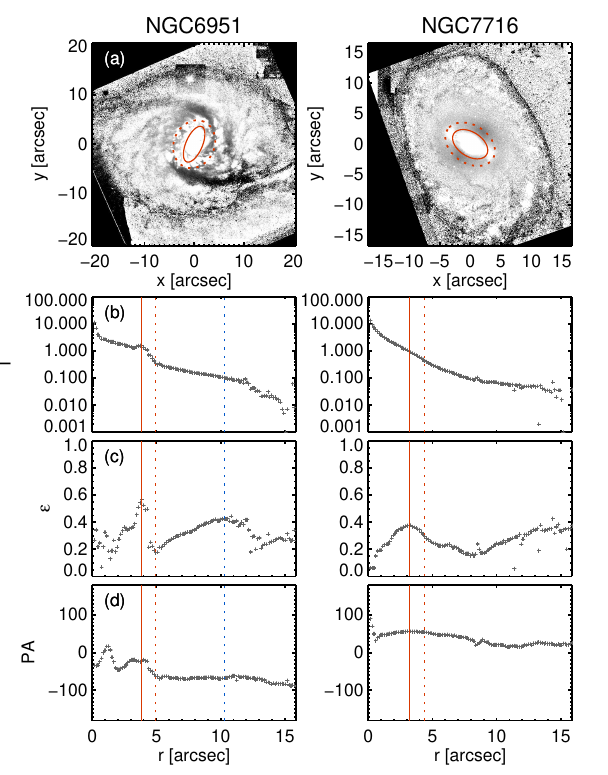}
\caption{Nuclear structures of NGC 6951 (left) and NGC 7716 (right) analyzed with BarIstA for ellipse fitting on HST/PC images. The top panel shows the unsharp masked images of the HST/PC images, with the nuclear ring in NGC 6951 and the nuclear bar in NGC 7716 outlined by red ellipses. The bottom three panels represent the radial profiles of intensity, ellipticity, and PA from top to bottom. Red solid lines indicate the sizes of the nuclear structures determined from the maximum ellipticity, which tend to underestimate the sizes compared to visual inspection. Red dotted lines indicate the size identified at the radius with minimum ellipticity for NGC 6951, and that determined through visual inspection for NGC 7716. These sizes are shown as ellipses in the top panels. An additional bar-like signature at $10.\!\arcsec3$ in NGC 6951 is indicated by the blue dotted lines, which will be discussed in Sections \ref{chap4} and \ref{chap5}.}
\label{Fig3}
\end{figure}

%NGC 6951 and NGC 7716 are classified as SAB(rs)bc and SAB(r)b galaxies, respectively, according to NED. Both galaxies are identified as hosting nuclear rings and included in the AINUR catalog \citep{2010Comeron}. 

We examine the nuclear features of the target galaxies using HST/WFPC F606W images, which offers high spatial resolution with a pixel scale of $0.\!\arcsec0996$ for WF (left panel) and $0.\!\arcsec0445$ for PC (right panel), as shown in Figure \ref{Fig2}. To highlight the nuclear structures, we apply unsharp mask processing (bottom panels), a technique that enhances detail and contrast by dividing the original image with its smoothed version \citep{1977Malin, 1983Malin, 2003Erwin, 2004Erwin}. This approach effectively suppresses the large-scale smoothed light distribution while amplifying faint and low-contrast features. The smoothed images were generated using $10\arcsec$-sized boxes for WF images and $5\arcsec$-sized boxes for PC images.

The nuclear ring of NGC 6951 is clearly resolved in both the PC image (Figure \ref{Fig2}(b)) and the unsharp masked image (Figure \ref{Fig2}(f)), consistent with previous studies \citep{1995Wozniak, 1996Friedli, 2006Knapen}. NGC 7716 displays a nuclear bar within the inner ring in the PC image (Figure \ref{Fig2}(d)), with this feature further accentuated in the unsharp masked image (Figure \ref{Fig2}(h)). This observation aligns with earlier studies that classified NGC 7716 as a double-barred galaxy \citep{2002Laine, 2004Erwin}. 

We further analyze nuclear structures using HST/PC images by fitting ellipses with BarIstA \citep{2019Lee, 2020Lee}, as shown in Figure \ref{Fig3}. The three bottom panels show the radial profiles of intensity, ellipticity, and PA from top to bottom. First, the sizes of the nuclear ring and nuclear bar are identified as the locations of maximum ellipticity in the ellipticity profile \citep{1995Martin, 1995Wozniak, 2004Jogee}, marked by red solid lines. The corresponding ellipses are overlaid with red solid lines on the HST/PC unsharp masked image in the top panel. In our target galaxies, the sizes of nuclear structures determined by the maximum ellipticity are likely underestimated compared to those estimated from visual inspection, as similarly found in previous studies \citep{2002Laurikainen, 2003Erwin, 2009Aguerri}. To mitigate this effect, we also indicate the sizes of the nuclear structures as the radius with minimum ellipticity for NGC 6951, and as the radius identified through visual inspection for NGC 7716. These are represented by red dotted lines. 

The sizes of the nuclear structures are then deprojected and summarized in Table \ref{Table2}. The deprojected radii can be estimated using the following relation: 
\begin{eqnarray}
r_{\rm deproj} = r_{\rm obs}\sqrt{\rm cos^2\phi+sin^2\phi/cos^2 \it i}
\end{eqnarray}
where $r_{\rm obs}$ is the observed radius, $\phi$ is the angle between the major axis of the nuclear structures and that of the galaxy disk, and $i$ is the inclination of the galaxy \citep{1995Martin, 2007Gadotti}. However, the values of $r_{\rm deproj}$ and $r_{\rm obs}$ for our target galaxies are nearly identical, as the nuclear structures are closely aligned with the major axes of their respective main disks.

Lastly, we note that NGC 7716 was selected from the AINUR catalog, which compiled galaxies with nuclear rings \citep{2010Comeron}. Historically, \citet{1980deVaucouleurs} identified NGC 7716 as hosting an inner ring. However, subsequent studies reclassified this structure as a nuclear ring due to the detection of a larger bar extending beyond the radius of the ring \citep{2006Knapen, 2008Mazzuca}. This reclassification is why NGC 7716 is included in the AINUR catalog \citep{2010Comeron}. 

However, considering the bar length of $33\arcsec$ reported in the AINUR catalog, the winding spiral arms, filled with a reddish hue and indicated by the white arrow in Figure \ref{Fig1}(b), could have been interpreted as the bar. The deprojection using the orientation parameters from the catalog further accentuates this feature. This serves as a representative example of how erroneous orientation parameters can significantly affect the measurement of bar length and strength, as discussed in Section \ref{chap2.2}. In contrast, the parameters we derived suggest that NGC 7716 hosts a weak bar ($18\arcsec$ in length, measured using the same method as in the AINUR catalog), surrounded by an inner ring, with a nuclear bar at its central region, as shown in Figure \ref{Fig2}(h). Further details will be discussed in Sections \ref{chap4} and \ref{chap5.2}.

\subsection{Nuclear Features from Spectroscopy} \label{chap3.2} %===================================

\begin{figure}
\centering
\includegraphics[width= 0.95\linewidth, trim = {10 280 290 30}, clip]{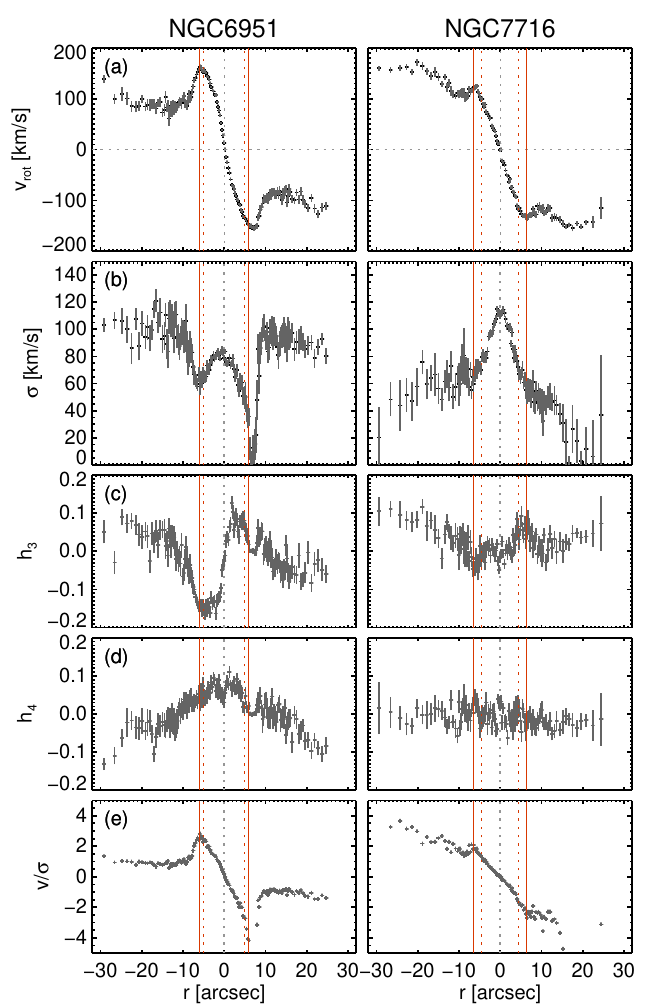}
\caption{Profiles of the stellar velocity ($v_{\rm rot}$), velocity dispersion ($\sigma$), the asymmetric ($h_3$) and symmetric ($h_4$) deviations from a Gaussian, and $v/\sigma$ for NGC 6951 (left) and NGC 7716 (right), arranged from top to bottom. All kinematic information is derived by applying pPXF to long-slit spectra aligned with the major axes of the disks. Uncertainties are estimated using Monte Carlo simulations with 100 iterations. The red solid lines represent the kinematic radius ($R_{\rm kin}$) of the nuclear disk, while the red dotted lines indicate the deprojected sizes ($r_{\rm deproj}$) of the nuclear ring in NGC 6951 and the nuclear bar in NGC 7716. The gray dotted lines mark the kinematic centers of galaxies.}
\label{Fig4}
\end{figure}

The LOSVDs can be modeled using Gauss-Hermite functions with four parameters: stellar velocity ($v$), velocity dispersion ($\sigma$), and the asymmetric ($h_3$) and symmetric ($h_4$) deviations from a Gaussian \citep{1993vanderMarel}. These parameters serve as powerful tools for understanding the stellar structures in galaxies \citep{2020Gadotti}. To derive these stellar kinematics, we applied the penalized-pixel fitting \citep[pPXF;][]{2004Cappellari, 2017Cappellari} to the sky-cleaned long-slit spectra obtained along the major axes of the disks. These spectra are the same as those used in the third panels of Figure \ref{FigA}, but are Voronoi-binned to achieve an S/N $\geq$ 20 \citep{2003Cappellari}. We used the wavelength range of 8500$-$8750~\text{\AA}, which covers the Ca II triplet lines at $z < 0.01$ for our target galaxies, while masking 8630$-$8670~\text{\AA} region to remove instrumental noise. The best-fit templates were constructed using the E-MILES library of stellar spectra \citep{2016Vazdekis}. Uncertainties were estimated using Monte Carlo simulations with 100 iterations.

Figure \ref{Fig4} presents $v_{\rm rot}$, $\sigma$, $h_3$, $h_4$, and $v/\sigma$ from top to bottom. The rotation curves (Figure \ref{Fig4}(a)) are derived from the observed stellar velocity as follows,
\begin{equation}
    v_{\rm rot}  = \frac{v_{\rm obs}-v_{\rm sys}}{\rm sin \it i}
\end{equation}
where $v_{\rm rot}$, $v_{\rm obs}$, and $v_{\rm sys}$ represent the rotation velocity, observed velocity, and systemic velocity, respectively, while $i$ denotes the inclination of the galaxy. $v_{\rm sys}$ was determined at the kinematic center, identified as the point where the velocity curve exhibits approximate symmetry between the approaching and receding sides, yielding 1414 $\pm~2~\rm km~s^{-1}$ for NGC 6951 and 2576 $\pm~2~\rm km~s^{-1}$ for NGC 7716. The heliocentric systemic velocities and distances of these galaxies are summarized in Table \ref{Table1}. The velocity dispersion (Figure \ref{Fig4}(b)) is determined after correcting for instrumental broadening during the pPXF process. The deprojected sizes of the nuclear structures are represented as red dotted vertical lines, as measured in Section \ref{chap3.1}.

The complex nuclear structures of our target galaxies are clearly evident across all kinematic profiles presented in Figure \ref{Fig4}. The top panels show that the $v_{\rm rot}$ profiles exhibit peaks around the nuclear structures, as indicated by the red vertical lines. These characteristics likely suggest the presence of dynamically decoupled components within the galaxies \citep{2001Emsellem}. Such components are typically identified by a sudden change in $v$, a drop in $\sigma$, an anticorrelation between $h_3$ and $v$, or elevated values of $h_4$ \citep{2001Emsellem, 2003Wozniak, 2004Chung, 2006Falcon-Barroso, 2020Gadotti}. These signatures of dynamically decoupled components are evident in both of our target galaxies.

\begin{figure}
\centering
\includegraphics[width= 0.85\linewidth, trim = {410 435 0 150}, clip]{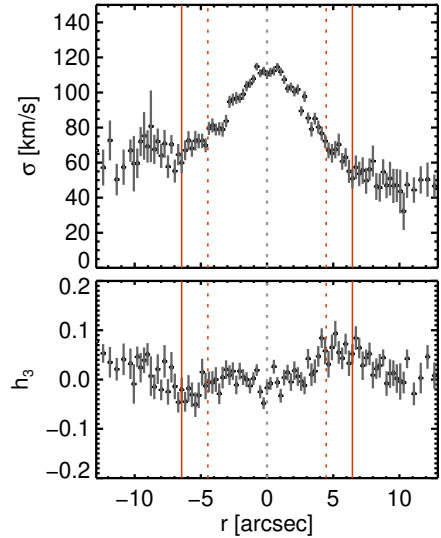}
\caption{Zoomed-in profiles of $\sigma$ (top) and $h_3$ (bottom) for NGC 7716. The $\sigma$ profile in the top panel appears flat in the central region, a characteristic of a $\sigma$-drop, while the $h_3$ profile in the bottom panel shows an anticorrelation with $v_{\rm rot}$ (right panel in Figure \ref{Fig4}(c). Uncertainties are estimated using Monte Carlo simulations with 100 iterations. The vertical lines correspond to those in Figure \ref{Fig4}.}
\label{Fig5}
\end{figure}

The left panel of Figure \ref{Fig4}(b) shows a broad decrease in the central region of $\sigma$ profile in NGC 6951, indicating the presence of a rotationally supported structure, a nuclear cold disk \citep{2003Wozniak, 2006Wozniak, 2020Gadotti}. While the central region of $\sigma$ profile in NGC 7716 (right panel) initially appears as a peaked nucleus, the zoomed-in profile in Figure \ref{Fig5} reveals a flat central region, characteristic of a $\sigma$-drop \citep{2014Mendez-Abreu}. A $\sigma$-drop typically appears as a flat or a dip in the central region of the $\sigma$ profile \citep{2001Emsellem, 2009Perez, 2014Mendez-Abreu}. Meanwhile, the upturn observed within the $\sigma$-drop in NGC 6951 may be attributed to kinematically hot spheroids or large-scale outflows driven by AGN activity \citep{2003Marquez, 2009Perez, 2020Bittner}. We note that NGC 6951 has been classified as either a Seyfert 2 galaxy \citep{1997Ho} or a Low-Ionization Nuclear Emission-line Region (LINER) galaxy \citep{2007Storchi}.

Additionally, near-circular orbits are characterized by an anticorrelation between $h_3$ and $v$, while elongated orbits exhibit a positive correlation between the two \citep{2004Chung, 2005Bureau, 2015Iannuzzi, 2020Gadotti}. In both galaxies, we observe an anticorrelation between $h_3$ (Figure \ref{Fig4}(c)) and $v_{\rm rot}$ (Figure \ref{Fig4}(a)), which further supports the presence of a nuclear stellar disk. A zoomed-in view of $h_3$ profile of NGC 7716 is also shown in Figure \ref{Fig5}. The small distortion in the central region of $h_3$ in NGC 7716 could be attributed to the presence of the nuclear bar.  In the case of NGC 6951, the elevated $h_4$ (Figure \ref{Fig4}(d)) further suggests the superposition of structures with distinct LOSVDs \citep{1994Bender, 2020Gadotti}.

As a result, the kinematic profiles of both target galaxies consistently indicate the presence of dynamically decoupled nuclear stellar disks. The $v/\sigma$ profile in Figure \ref{Fig4}(e) provides the kinematic radius of the nuclear disk ($R_{\rm kin}$), where the profile peaks due to the rapidly rotating disk \citep{2020Gadotti}. We present $R_{\rm kin}$ as red vertical solid lines across all profiles in Figures \ref{Fig4} and \ref{Fig5}, and summarize both $R_{\rm kin}$ and the corresponding $v/\sigma$ values at $R_{\rm kin}$ in Table \ref{Table2}. In our observations, both a nuclear ring and a nuclear bar, as indicated by red dotted lines, are located within $R_{\rm kin}$. Further discussion on the nuclear disk will be provided in Sections \ref{chap5.1} and \ref{chap5.2}.

\section{Bar Properties} \label{chap4} %===================================
Bars are typically characterized by three main parameters: bar length, strength, and pattern speed \citep{2015Aguerri}, all of which are crucial for studying bar evolution. In this study, we measure the bar length and strength using $i-$band images obtained from the PS1 archive, and calculate the bar pattern speed using the Gemini/GMOS long-slit spectra described in Section \ref{chap2.3}.

\subsection{Bar Length and Strength} \label{chap4.1} %======================

\begin{figure*}
\centering
\includegraphics[width= 0.9 \linewidth, trim = {20 430 190 30}, clip]{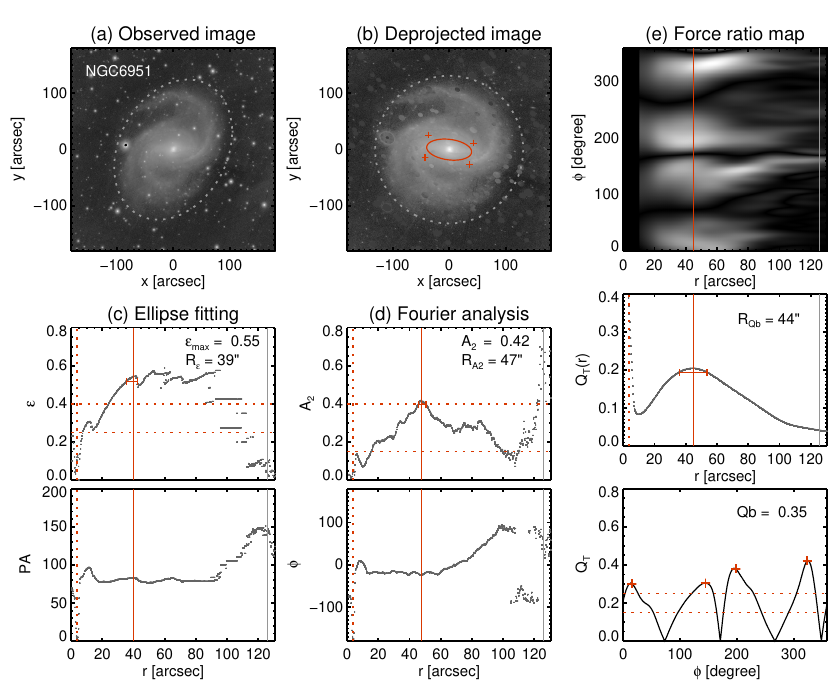}
\caption{Procedure of measuring the bar length and strength of NGC 6951 using the three main methods with BarIstA. (a) $i-$band observed image from PS1, (b) deprojected image using orientation parameters, (c) ellipse fitting method: ellipticity (top) and PA (bottom) profiles, (d) Fourier analysis: $A_2$ (top) and $\phi$ (bottom) profiles, (e) force ratio map method: force ratio map in polar coordinates ($r, \phi$) (top), $Q_T$ radial profile (middle), and azimuthal profile at $R_{Qb}$ (bottom). Disk shapes based on orientation parameters are overlaid by gray dotted lines in (a) and (b). The measured bar lengths and strengths are listed in relevant panels. Red solid lines indicate the bar length, while red horizontal dotted lines represent the criteria for defining weakly and strongly barred galaxies, respectively. Four red crosses mark the four corners of the bar structure in (b) and the bottom panel of (e). Errors in the bar lengths are indicated by horizontal error bars in the middle row. Red vertical dotted lines indicate the deprojected size of the nuclear ring, as measured through ellipse fitting from the HST/PC image in Section \ref{chap3}. $R_{25}$ is indicated by gray vertical solid lines.}
\label{Fig6}
\end{figure*}

\begin{figure*}
\centering
\includegraphics[width= 0.9 \linewidth, trim = {20 60 190 390}, clip]{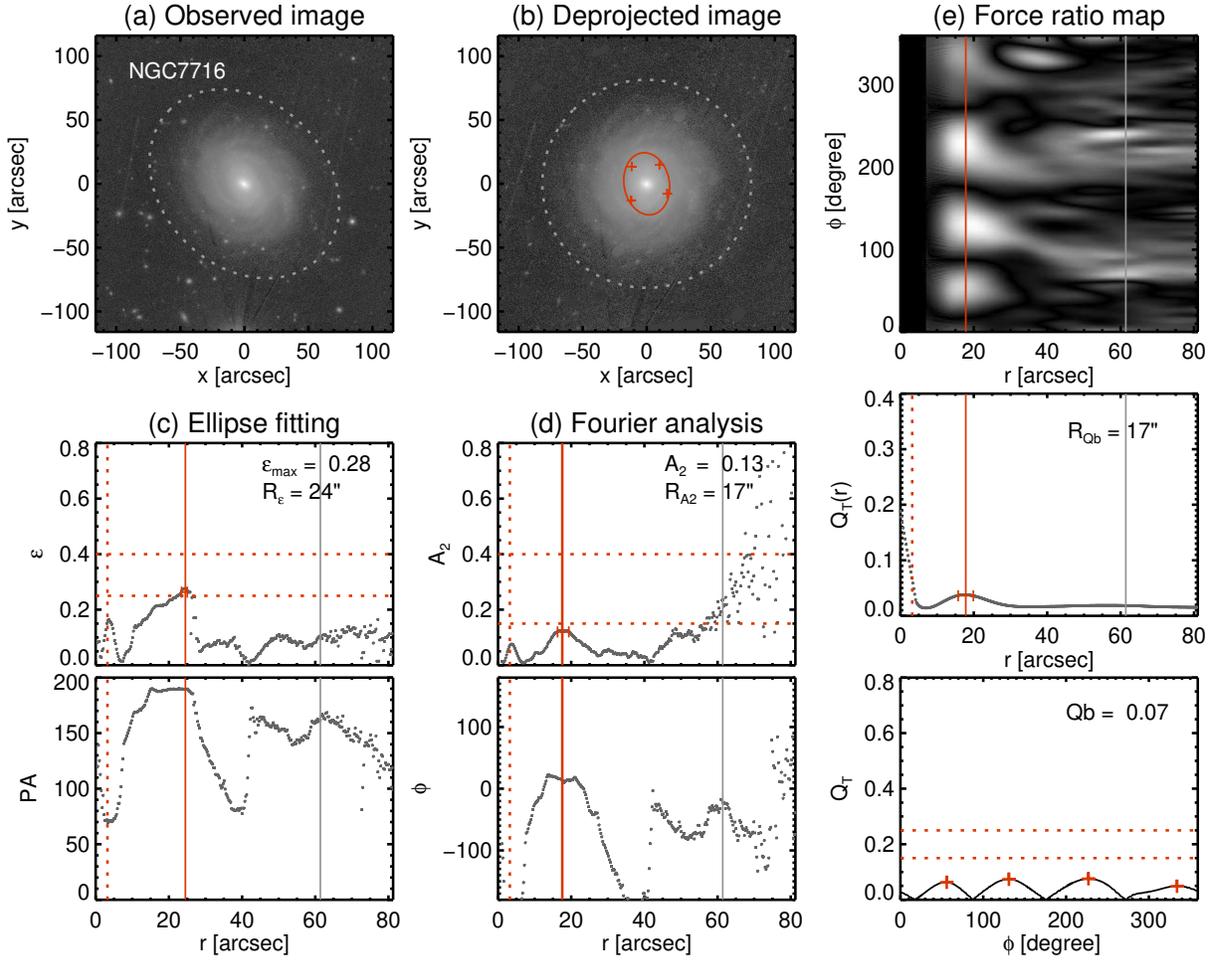}
\caption{Similar to Figure \ref{Fig6}, but for NGC 7716. The nuclear bar identified in the HST/PC image in Section \ref{chap3} and indicated by red vertical dotted lines, is also detected through ellipse fitting (c) and Fourier analysis (d).}
\label{Fig7}
\end{figure*}

Bar length and strength were initially estimated through visual inspection \citep{1973Nilson, 1991deVaucouleurs, 2015Ann, 2015Buta}, a practice still widely used thanks to contributions from citizen scientists in the Galaxy Zoo project \citep{2008Lintott, 2011Masters, 2014Simmons, 2025Geron}. More recently, this visually driven approach has evolved into a scalable framework through machine learning trained on Galaxy Zoo classifications. Machine learning methods have shown remarkable performance in processing large-scale survey data, including those from the Dark Energy Spectroscopic Instrument (DESI; \citealp{2023Walmsley, 2023Geron}).
%and the James Webb Space Telescope (JWST; \citealp{2025Geron}).

On the other hand, automated techniques have also been developed to analyze bar length and strength, enabling quantitative measurements and serving as powerful tools for detailed investigation of bar structures. To this end, we adopted three widely used methods: ellipse fitting \citep{1995Martin, 1995Wozniak, 2004Jogee}, Fourier analysis \citep{1990Ohta, 2002Laurikainen, 2004Laurikainen}, and the force ratio map \citep{2001Buta, 2002Lauri_Salo, 2016Diaz, 2020Lee}. The bar lengths obtained from these methods are defined as $R_\epsilon$, $R_{A_2}$, and $R_{Q_b}$, respectively, while the bar strengths are characterized by $\epsilon_{\rm max}$, $A_2$, and $Q_b$. Ellipse fitting and Fourier analysis are described in detail in \citet{2019Lee}, while the force ratio map method is explained in \citet{2020Lee}.

We applied all three methods using BarIstA \citep{2019Lee, 2020Lee} on sky-subtracted PS1 $i$-band images. Figures \ref{Fig6} and \ref{Fig7} illustrate the procedure for measuring bar length and strength. The observed image (panel (a)) was masked and deprojected based on the orientation parameters listed in Table \ref{Table1}, as outlined by the gray dotted ellipses. The resulting image is shown in panel (b), where the disk outline has been transformed into a perfect circle. The red ellipse traces the bar identified through ellipse fitting, while the four crosses mark the four corners of the bar structure determined using the force ratio map, as detailed below. 

%These measurements are typically derived from deprojected images \citep{2002Laurikainen, 2002Laine, 2019Lee, 2020Lee}. \

The results from ellipse fitting (column (c)), Fourier analysis (column (d)), and the force ratio map (column (e)) are presented from left to right. In general, bar strengths ($\epsilon_{\rm max}$ and $A_2$) are defined by the maximum values in the $\epsilon$ and $A_2$ profiles \citep{1995Martin, 2002Laurikainen, 2002Lauri_Salo}, as shown in the middle row. Bar lengths ($R_{\epsilon}$, $R_{A2}$, and $R_{Qb}$) are determined by the radii where the bar strengths are measured. The PA and the phase ($\phi$) are typically constant across the bar structures, remaining within $\pm$10$^\circ$ \citep{2009Aguerri, 2009Marinova, 2019Lee}, as shown in the bottom row. All red solid lines in Figure \ref{Fig6} depict the measured bar lengths. For NGC 6951, however, $\epsilon_{\rm max}$ appears along a spiral arm beyond $\sim50\arcsec$, aligned with the bar PA. Therefore, we adopted the local maximum near $\sim40\arcsec$ to determine the bar length and strength.

In Fourier analysis, $A_2$ is defined as follows,
\begin{equation}
A_{\rm 2} = \rm max\left(\frac{\sqrt{{\it a}_2^2+{\it b}_2^2}}{{\it a}_0}\right)    
\end{equation}
where $a_0$, $a_2$, and $b_2$ are the Fourier coefficients \citep{2013Athanassoula, 2019Seo, 2022Lee}. We note that this value is half of that used in some previous studies, based on their definition \citep{1990Ohta, 2002Laurikainen, 2019Lee, 2020Lee}.
%These measurements are robust and reliable, though they do not represent the full length of the bar \citep{1995Wozniak, 2002Laurikainen, 2002Lauri_Salo, 2016Diaz}.

The rightmost panels (column (e)) illustrate the procedure of analyzing the force ratio map, which represents the transverse-to-radial force derived from the potential map \citep{1981Combes, 2001Buta, 2020Lee}. When displayed in polar coordinates ($r, \phi$), as shown in the top panel, bar structures appear as four horizontal thick slabs, aligned with similar length ($r$) and width ($\phi$) \citep{2020Lee}. The force ratio is defined as follows,
\begin{eqnarray}\label{eq3}
Q_{\rm T}(r,\phi) = \frac{F_{\rm T}(r,\phi)}{\langle F_{\rm R}(r)\rangle}, 
\end{eqnarray}
where the radial force is described by 
\begin{eqnarray}\label{eq5}
\langle F_{\rm R}(r) \rangle \equiv \frac{d\Phi_{0}(r)}{dr},
\end{eqnarray}
and the transverse force is given by
\begin{eqnarray}\label{eq6}
F_{\rm T}(r,\phi) \equiv \left|\frac{1}{r}\frac{\partial\Phi(r,\phi)}{\partial\phi}\right|.
\end{eqnarray}
Here, $\Phi_0$ represents the $m=0$ Fourier component of the gravitational potential \citep{1981Combes, 2001Buta}. 

%%---- Table2 ---------------
\begin{deluxetable*}{clccccccccccc}
 \tablecaption{Measurements of the galactic structures \label{Table2}}
 \tablehead{
 \colhead{Galaxy} &
 \colhead{Unit} &
 \multicolumn{2}{c}{Nuclear substructure} &
 \multicolumn{2}{c}{Nuclear Disk} &
 \multicolumn{3}{c}{Bar length} &
 \multicolumn{3}{c}{Bar strength} &
 \colhead{Disk} \\
\colhead{ } &
\colhead{ } &
 \multicolumn{2}{c}{($R_{\rm NR}$, $R_{\rm NB}$)} &
 \multicolumn{2}{c}{($R_{\rm ND}$)} &
 \multicolumn{3}{c}{($R_{\rm bar}$)} &
 \multicolumn{3}{c}{($S_{\rm bar}$)} &
 \colhead{ } \\
% \colhead{} &
% \multicolumn{3}{c}{(arcsec)} &
% \colhead{} &
% \multicolumn{3}{c}{(arcsec)} &
% \multicolumn{3}{c}{} &
% \colhead{(arcsec)}\\
 \colhead{} &
  \colhead{} &
\colhead{$R_{\epsilon}$} &
\colhead{$R_{\rm visual}$} &
\colhead{$R_{\rm kin}$} &
\colhead{$v/\sigma$} &
 \colhead{$R_{\epsilon}$} &
 \colhead{$R_{A2}$} &
 \colhead{$R_{Qb}$} &
 \colhead{$\epsilon_{\rm max}$} &
 \colhead{$A_2$} &
 \colhead{$Q_b$} &
 \colhead{$R_{25}$} \\
}
\startdata
NGC 6951 & arcsec & 3.9 & 5.0 & 5.9 & 2.8 & 40.0$^{+2.6}_{-4.6}$ & 47.7$^{+2.8}_{-2.1}$ & 44.9$^{+8.8}_{-8.8}$ & 0.55 & 0.42 & 0.36 & 126.7 \\
         & kpc & 0.39 & 0.49 & 0.59 & - & 3.92$^{+0.06}_{-0.12}$ & 4.77$^{+0.07}_{-0.05}$ & 4.49$^{+0.22}_{-0.22}$ & - & - & - & 12.66 \\
NGC 7716 & arcsec & 3.2 & 4.5 & 6.5 & 2.1 & 24.5$^{+0.5}_{-1.0}$ & 17.5$^{+1.8}_{-1.3}$ & 17.8$^{+2.1}_{-2.1}$ & 0.28 & 0.13 & 0.07 & 61.4\\ 
        & kpc & 0.58 & 0.81 & 1.16 & - & 4.30$^{+0.02}_{-0.05}$ & 3.18$^{+0.08}_{-0.06}$ & 3.22$^{+0.09}_{-0.09}$ & - & - & - & 11.12 \\
\enddata
\tablecomments{\footnotesize The sizes of the structures are represented by $R_{\rm NR}$, $R_{\rm NB}$, $R_{\rm ND}$, and $R_{\rm bar}$, corresponding to a nuclear ring, nuclear bar, nuclear disk, and primary bar, respectively. The subscripts in the third row indicate the measurement methods used. The nuclear structures of NGC 6951 and NGC 7716 are a nuclear ring (NR) and a nuclear bar (NB), respectively.}
\end{deluxetable*}
%\vspace{-5mm} % 간격을 5mm 줄임
%%---------------------------

The force ratio map (top panel) can be analyzed in both the radial (middle panel) and azimuthal (bottom panel) directions. The radial profile (middle panel) of the force ratio map (top panel), $Q_{T}(r)$, is obtained by averaging $Q_T$ over all azimuthal angles at each radius. This profile initially decreases in the bulge-dominated region and then rises to a maximum, which defines the bar length ($R_{Qb}$), marked by red solid vertical lines. The azimuthal profile (bottom panel), on the other hand, is constructed by plotting the $Q_T$ values across azimuthal angles at a fixed radius. Notably, when the azimuthal profile is examined at $R_{Qb}$, it reveals four distinct peaks, as shown in the bottom panel. These four peaks correspond to the four corners of the bar, which also align with the four slabs in the force ratio maps (top panel) and are marked as red crosses on the deprojected images (panel (b)). The bar strength ($Q_b$) is calculated as the mean $Q_T$ value at these four peaks \citep{2020Lee}. All measured bar length ($R_{\epsilon}$, $R_{A2}$, and $R_{Qb}$) and strength ($\epsilon_{\rm max}$, $A_2$, and $Q_b$) are provided in the relevant panels of Figures \ref{Fig6} and \ref{Fig7}, as well as in Table \ref{Table2}. Errors in the bar lengths are estimated from the widths of the peaks in the $\epsilon$, $A_2$, and $Q_T(r)$ profiles, determined at the radii where $\epsilon$, $A_2$, and $Q_T$ reach 95\% of their peak values,  following the approach of \citet{2021Cuomo}. For comparison, the deprojected sizes of the nuclear structures (red dotted vertical lines) and the galaxy disk, measured at 25 $\rm mag/arcsec^2$ ($R_{25}$, gray solid vertical lines), are also included.

The three methods described above for measuring bar length and strength are useful for identifying barred galaxies \citep{2019Lee, 2020Lee}. While additional characteristics, such as the transition between a bar and a disk, and a constant PA or phase, should also be considered \citep{1995Wozniak, 2002Laurikainen, 2004Jogee, 2007Menendez-Delmestre, 2007Marinova, 2009Aguerri, 2019Lee, 2020Lee}, each method provides specific criteria for bar strength that aid in identifying barred galaxies. These criteria can be summarized as $\epsilon_{\rm max} \geq 0.25$, $A_2 \geq 0.15$, and $Q_b \geq 0.15$ \citep{2002Laurikainen, 2007Marinova, 2008Barazza, 2009Marinova, 2019Lee, 2020Lee}. While \citet{2002Laurikainen} suggested a criterion of $I_2/I_0 > 0.3$, this can be interpreted as $A_2 > 0.15$, since $I_2/I_0$ is larger than $A_2$ by a factor of two. We overlay these bar strength criteria as red horizontal dotted lines in the relevant panels of Figures \ref{Fig6} and \ref{Fig7}. Additionally, we plot criteria of $\epsilon_{\rm max} \geq 0.4$, $A_2 \geq 0.4$, and $Q_b \geq 0.25$ \citep{2004Jogee, 2019Cuomo, 2019Lee, 2020Lee}, which have been used to indicate strongly barred galaxies. 

Therefore, given the criteria for the bar strength, NGC 6951 can be classified as strongly barred, with $\epsilon_{\rm max} = 0.55$, $A_2 = 0.42$, and $Q_b = 0.36$. On the other hand, NGC 7716 only barely meets the requirement for barred galaxies on the criterion of $\epsilon_{\rm max} = 0.25$ and rarely satisfies the criterion in Fourier analysis. Moreover, the results from the force ratio map do not support the presence of a bar. Specifically, the four peaks in the azimuthal profile of the force ratio map (bottom rightmost panel in Figure \ref{Fig7}) are significantly low. Therefore, it is difficult to conclude that the oval feature surrounded by the inner ring of NGC 7716 represents a weak bar. We note that both of our target galaxies, NGC 6951 and NGC 7716, have been visually classified as SAB in the RC3 catalog \citep{1991deVaucouleurs}, as summarized in Table \ref{Table1}.

%This suggests that although the oval feature  NGC 7716 appears bar-like in shape, it might be sparsely distributed with a low density when considering the bar signatures derived from the potential map. 

%When distinguishing between strongly barred galaxies (SB) and weakly barred galaxies (SAB) through visual inspection, the shape of the bar, relative bar length, and contrast are commonly used as criteria \citep{2015Ann, 2015Buta, 2019Lee}. While SABs are characterized by their oval shape, shorter bars, and less prominent features compared to SBs, distinguishing SABs remains challenging, not only from SBs but also from non-barred galaxies (SAs) \citep{2019Lee, 2021Cuomo, 2023Geron}. Notably, both of our target galaxies, NGC 6951 and NGC 7716, have been visually classified as SAB in the RC3 catalog \citep{1991deVaucouleurs}, as summarized in Table \ref{Table1}.

Bar length tends to depend on the properties of host galaxies, such as stellar mass or galaxy size \citep{2019Erwin, 2024Erwin, 2019Cuomo, 2021Kim, 2023Aguerri}, which differs from bar strength \citep{2016Diaz, 2020Cuomo, 2022Lee}. In particular, bar length is tightly correlated with galaxy size, including scale length ($h_r$), effective radius ($R_{\rm eff}$), as well as the radii enclosing 50$\%$ ($R_{50}$) and 90$\%$ ($R_{90}$) of the total galaxy luminosity \citep{1987Ann, 2021Kim}. As a result, bar length is often examined as a relative size with respect to galaxy size, such as $h_r$ or $R_{25}$ \citep{1985Elmegreen, 1995Martin, 2007Menendez-Delmestre, 2009Aguerri, 2020Lee}. For our target galaxies, when investigating $R_{\rm bar}/R_{25}$, the bar size of NGC 6951 ranges from 0.32 to 0.38, while NGC 7716 ranges from 0.29 to 0.40, depending on the methods used. This will be discussed further in Section 5.3.

%\vspace{-5mm} % 간격을 5mm 줄임
\subsection{Bar Pattern Speed} \label{chap4.2} %=======================

\begin{figure*}
\centering
\includegraphics[width= \linewidth, trim = {0 350 15 30}, clip]{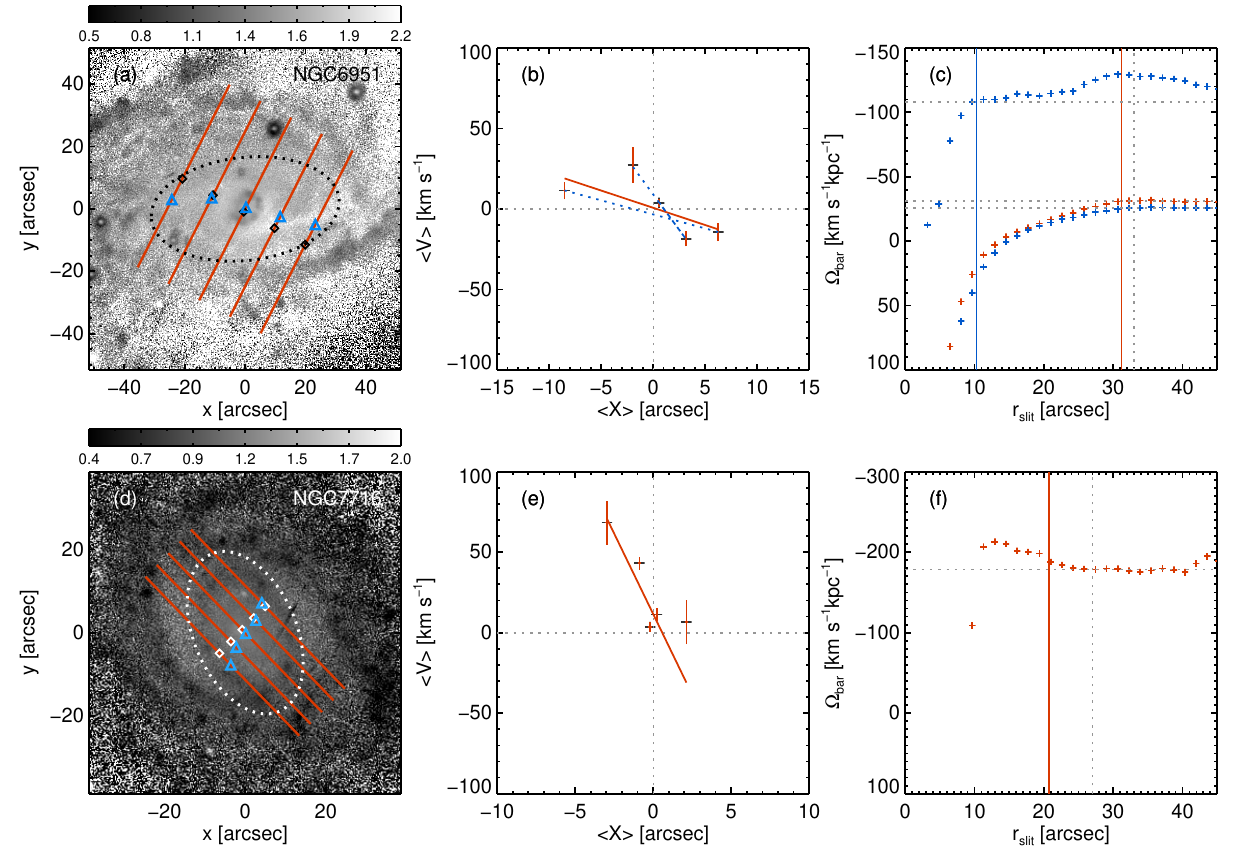}
\caption{Luminosity-weighted positions, shown as cyan triangles, are overlaid on the slits indicated by red solid lines, while black (top) or white (bottom) squares mark the midpoints of the slit (leftmost). $g\!-\!i$ color maps serve as the background with dotted ellipses outlining the bar structure. The corresponding color bars are displayed above each image. The luminosity-weighted velocity is plotted as a function of the luminosity-weighted position (middle). Each slit yields one data point on this plane, with uncertainties derived from
chi-square minimization. The slope of the straight fitting line, indicated by the red solid lines, gives $\rm sin~\it i~\rm \Omega_{\rm bar}$.  The blue dotted lines in NGC 6951 (b) show the results fitted with two different pattern speeds. The gray dotted lines serve as a guide, showing zero values of $\langle X \rangle$ and $\langle V \rangle$. The rightmost column shows $\rm \Omega_{\rm bar}$ as a function of slit length. Red plus symbols represent $\Omega_{\rm bar}$ derived assuming a single-component fit, while blue plus symbols indicate the results from a two-component fit. The red solid vertical lines mark the bar length, measured from observed images using ellipse fitting. The gray dotted vertical lines indicate the radius where the S/N exceeds 3. The results at these radii are shown in the leftmost and middle panels. In panel (c) for NGC 6951, the blue solid vertical line represents the radius corresponding to an additional bar-like feature, as identified in the $\epsilon$ and PA profiles from HST/PC photometry, shown in the left panels of Figures \ref{Fig3}(c) and (d).}
\label{Fig8}
\end{figure*}

%%---- Table3 ---------------
\begin{deluxetable*}{clccc}
 \tablecaption{Bar pattern speeds of sample galaxies
 \label{Table3}}
 \tablehead{
 \colhead{Galaxy} &
 \colhead{Unit} &
 \colhead{One component} &
 \multicolumn{2}{c}{Two components} \\
 \colhead{} &
 \colhead{} &
 \colhead{$\rm \Omega_{\rm bar}$} &
 \colhead{$\rm \Omega_{\rm bar}$} &
 \colhead{$\rm \Omega_{\rm NS}^1$}  \\
}
\startdata
NGC 6951 &  $\rm km~s^{-1}arcsec^{-1}$ & $-3.2\pm0.7$ & $-2.5\pm0.8$ & $-10.9\pm6.4$  \\
         & $\rm km~s^{-1}kpc^{-1}$ & $-31.9\pm7.1$ & $-25.5\pm7.8$ & $-108.7\pm63.5$ \\
NGC 7716$^2$ & $\rm km~s^{-1}arcsec^{-1}$ & $-32.1\pm4.8$ & - & - \\
         & $\rm km~s^{-1}kpc^{-1}$ & $-177.3\pm26.6$ & - & - \\ 
\enddata
\tablecomments{\footnotesize $^1$$\rm \Omega_{\rm NS}$ is measured at $r\sim10$, where it begins to converge. \\
$^2$Although $\Omega$ can technically be measured in NGC 7716, it may be influenced by external factors and may not represent the actual $\Omega_{\rm bar}$. }
\end{deluxetable*}
%\vspace{-5mm} % 간격을 5mm 줄임
%--------------------------

Bar pattern speed, $\Omega_{\rm bar}$ can be directly obtained from multiple long-slit observations using the TW method \citep{1984Tremaine, 1995Merrifield}, which is expressed as follows: 
\begin{align}\label{eq7}
\Omega_{\rm bar} & = \frac{1}{\rm sin~\it i}\frac{\int_{-\infty}^{\infty} l(x)[\bar v_{\rm obs}(x)-v_{\rm sys}]dx}{\int_{-\infty}^{\infty} l(x)(x-x_0)dx} \\
%& = \frac{1}{\rm sin~\it i}\frac{\langle\bar v_{\rm obs}\rangle-v_{\rm sys}}{\langle x \rangle-x_0} \\
& = \frac{1}{\rm sin~\it i}\frac{\langle V \rangle}{\langle X \rangle}
\end{align}
where $i$, $v_{\rm sys}$, and $x_0$ denote the inclination, systemic velocity, and center of the galaxy, respectively. The functions $l(x)$ and $\bar v_{\rm obs}(x)$ represent the mean luminosity and observed LOS velocity of stars at position $x$ along the slit, respectively \citep{1984Tremaine, 1995Merrifield}. The luminosity profile $l(x)$ was derived from spectra collapsed along the wavelength axis. Consequently, $\langle X \rangle$ and $\langle V \rangle$ correspond to the luminosity-weighted position and velocity along each slit. In practice, each slit yields a pair of $\langle X \rangle$ and $\langle V \rangle$, and a linear fit to these pairs across multiple slits directly gives $\rm sin~\it i~\rm \Omega_{\rm bar}$, as shown in the middle panels of Figure \ref{Fig8}.

Figure \ref{Fig8} illustrates the procedure for measuring $\Omega_{\rm bar}$ in our target galaxies. The leftmost columns show the positions of five long-slits, marked with red solid lines, aligned with the major axis of the disk. Luminosity-weighted positions, $\langle X \rangle$, are shown as cyan triangles on each slit, while black or white squares mark the slit midpoints. In the background, the $g\!-\!i$ color maps highlight bluer features such as nuclear rings, inner rings, or spiral arms. The bar shapes are outlined with dotted ellipses in panels (a) and (d). 

For NGC 6951, in panel (a), the five $\langle X \rangle$ positions closely follow the major axis of the bar. Notably, they align with the dust lane, which appears as a lighter-colored region. In contrast, panel (d) for NGC 7716 shows that $\langle X \rangle$ positions do not align with the major axis of the bar. This discrepancy may indicate that the contribution of the bar to the luminosity along each slit is not significant in NGC 7716. This may be partly due to the small angular offset between the major axis of the bar and that of the disk. 

The luminosity-weighted velocity, $\langle V \rangle$, was directly derived by measuring the velocity from spectra collapsed along the spatial direction, which effectively weights the luminosity, as pixels with higher signal contribute more to the resulting one-dimensional spectrum \citep{1995Merrifield, 2007Corsini, 2015Aguerri, 2019aCuomo}. This method has the advantage of yielding higher S/N measurements compared to integrating velocities measured at each pixel, weighted by the corresponding luminosity \citep{2019Zou}. However, when collapsing over $\sim$ 400 pixels, we found that instrumental noise also accumulates. Although the noise per pixel is negligible, its combined effect can sometimes hinder the pPXF fitting. Therefore, we measured the velocity by fitting Gaussian profiles to the Ca II triplet lines, with uncertainties estimated through chi-square minimization.

Finally, the five long-slit observations yield five data points for $\langle X \rangle$ and $\langle V \rangle$, as shown in the middle column of Figure \ref{Fig8}. The slope of the best-fitting straight line, shown in red solid line, corresponds to $\rm sin~\it i~\rm \Omega_{\rm bar}$. Using the galaxy inclination listed in Table \ref{Table1}, $\Omega_{\rm bar}$ is calculated as $-3.2\pm0.7~\rm km~s^{-1}~arcsec^{-1}$ for NGC 6951 and $-32.1\pm4.8~\rm km~s^{-1}~arcsec^{-1}$ for NGC 7716, which are summarized in Table \ref{Table3}. Uncertainties were estimated using Monte Carlo simulations with 1000 iterations. Errors in $\langle X \rangle$ are not considered, as they are negligible \citep{2015Aguerri, 2019Guo}.

As described so far, this method is direct and straightforward. However, asymmetries caused by other structures, such as spiral arms or lopsideness, could influence the measurements. It is therefore crucial to confirm that the contributions to the integrals primarily originate from the bar and not from other structures or artifacts \citep{2019Zou, 2019aCuomo}. The positions of $\langle X \rangle$ in the leftmost column of Figure \ref{Fig8} demonstrate that $\langle X \rangle$ for NGC 6951 effectively reflects the contribution of the bar, whereas for NGC 7716, the bar contributes minimally to the luminosity integrals. In the middle column, the data points deviate from the straight-line fit, differing from the typical trend observed in $\Omega_{\rm bar}$ measurements \citep{1995Merrifield, 2019aCuomo, 2024Cuomo}. Therefore, further investigation is required, which will be discussed in Sections \ref{chap4.2.1} and \ref{chap4.2.2}.

In addition, various factors that could influence the measurement of $\Omega_{\rm bar}$ have been examined, including the alignment of slits ($\rm PA_{slit}$), slit length, slit width, number of slits, wavelength range, and others \citep{2003Debattista, 2019Zou, 2019Guo, 2019aCuomo, 2024Cuomo, 2020Garma-Oehmichen}. Among these, the misalignment between the slit position ($\rm PA_{slit}$) and $\rm PA_{disk}$ introduces the most significant uncertainties \citep{2003Debattista, 2019Zou, 2020Garma-Oehmichen}. In such cases, the axisymmetric light from the disk fails to cancel out, amplifying the error \citep{2020Garma-Oehmichen}. Thus, obtaining an accurate measurement of $\rm PA_{disk}$ is essential. 
%\textcolor{orange}{with deviations as small as $5^\circ$ resulting in uncertainties of approximately $50\%$ 
%\textcolor{orange}{This will be further discussed in Section \ref{chap6.1}.}

Regarding the slit length for spatial coverage, the integrals in Equation \ref{eq7} theoretically extend from $-\infty$ to $\infty$. However, in practice, a low S/N in the outer regions can introduce false signals \citep{2019Guo}, and beyond the bar, potential contributions from spiral arms must also be considered. Fortunately, $\Omega_{\rm bar}$ generally converges to a reliable value when the slit length slightly exceeds the bar length, $\rm R_{bar}$ \citep{2019Guo, 2019Zou, 2020Garma-Oehmichen}. Therefore, we assess the convergence of $\Omega_{\rm bar}$ as a function of slit length, as shown in the rightmost column of Figure \ref{Fig8}. The red plus symbols show the variation of $\Omega_{\rm bar}$ with increasing slit length, while blue plus symbols represent the case where $\Omega_{\rm bar}$ is analyzed with two components, as will be discussed in Section \ref{chap4.2.1}. The bar length, $\rm R_{bar}$, measured from the observed images in the leftmost panels, is indicated by red solid vertical lines, while gray dotted solid lines indicate the slit length at which an S/N $>$ 3 per pixel is achieved. We find that $\Omega_{\rm bar}$ begins to converge when the slit length approaches $\rm R_{bar}$ in both NGC 6951 and NGC 7716. The values of $\Omega_{\rm bar}$ remain stable to some extent, even when the S/N at certain pixels does not reach 3. The leftmost and middle columns correspond to cases where the slit length ensures S/N $>$ 3 per pixel. Lastly, factors such as slit width, the number of slits, and the wavelength range generally do not introduce significant differences in $\Omega_{\rm bar}$ measurements \citep{2019Zou, 2019aCuomo, 2024Cuomo}.
%\textcolor{orange}{Further examination will be provided in Section \ref{chap6.1}.}
% The exact convergence length differs among galaxies \citep{2019Zou, 2020Garma-Oehmichen}. Therefore, selecting a slit length within the optimal range where $\Omega_{\rm bar}$ converges ensures a reliable estimate \citep{2019Zou, 2019aCuomo}.

\subsubsection{Two Pattern Speeds in NGC 6951} \label{chap4.2.1}

The five data points from the long-slit observations of NGC 6951, shown in Figure \ref{Fig8}(b), exhibit significant deviations from the single straight fitting line indicated in red. Instead, they are better represented by two separate straight lines, shown as blue dotted lines, suggesting the presence of two distinct pattern speeds \citep{2003Corsini, 2006Maciejewski, 2009Meidt}. This feature of the TW integrals resembles the results for NGC 2950, where \citet{2003Corsini} first identified two pattern speeds using the TW method, attributed to a rapidly rotating nuclear bar and a slowly rotating primary bar. 

We also examine the convergence of $\Omega_{\rm bar}$ for the two components as a function of slit length, shown in blue, in panel (c) of Figure \ref{Fig8}. The slower component of $\Omega_{\rm bar}$ converges when the slit length is approximately $R_{\rm bar}$, whereas the faster component stabilizes more quickly, around 10$\arcsec$ of the slit length. Interestingly, this radius coincides with a bar-like feature in the $\epsilon$ and PA profiles from HST/PC photometry, as indicated by the blue dotted line in the left panels of Figures  \ref{Fig3}(c) and (d). This feature is characterized by an increasing $\epsilon$ ($\epsilon_{\rm max} = 0.43$ at $r \sim 10.\!\arcsec3$) and a nearly constant PA. Similar bar-like signatures are observed at a comparable radius in ellipse fitting and Fourier analysis using lower resolution PS1 photometry (Figure \ref{Fig6}(c) and (d)). However, neither the PC image nor the corresponding unsharp masked image (Figures \ref{Fig2}(b) and (f)) provides clear evidence of an additional bar-like structure at this radius. Two pattern speeds of NGC 6951 are summarized in Table \ref{Table3} and will be discussed further in Section \ref{chap5.3}.

%While NGC 6951 does not host a nuclear bar, \textcolor{orange}{the nuclear spiral arms starting from the nuclear ring, shown in the top left unsharp mask in Figure \ref{Fig3}}, may have their own pattern speed \textcolor{orange}{when considering the slit and $\langle X \rangle$ positions.} 

%\subsubsection{Other Asymmetric Sources in NGC 7716} \label{chap4.2.2}
\subsubsection{Uncertain Bar Pattern Speed of NGC 7716}\label{chap4.2.2}

For NGC 7716, shown in Figure \ref{Fig8}(e), the three rightmost data points are close to zero in $\langle V \rangle$, indicating that there are no dominant non-axisymmetric components \citep{2019Zou}. This result aligns with the discrepancy between $\langle X \rangle$ and the bar major axis in Figure \ref{Fig8}(d), as well as with the classification of NGC 7716 as a non-barred galaxy based on Fourier analysis and the force ratio map, discussed in Section \ref{chap4.1}. 

However, the two leftmost data points exhibit a significant redshift compared to the others. We can determine whether these are genuine features or artifacts caused by noise in a few pixels by examining the velocity profiles along each slit, as shown in Figure \ref{FigA}. The left column of Figure \ref{FigA}, for NGC 6951, shows that the velocities in all slits are approximately symmetric with respect to the systemic velocity, as indicated by the red solid horizontal lines. In contrast, the top two panels in the right column exhibit prominent redshifts, corresponding to the two leftmost slits of NGC 7716 in Figure \ref{Fig8}(d). This confirms that the redshifted data points in Figure \ref{Fig8}(e) represent genuine features rather than noise. 

Therefore, the kinematics of NGC 7716 appear more complex than anticipated, although the origin of the redshifted component remains unclear. We find a hint in a deep image from SDSS Stripe 82, which reveals tidal streams around NGC 7716 with an $r$-band surface brightness of $\sim 27 \rm ~mag/arcsec^2$ (see Figure 7 in \citealt{2018Bakos}; see also DESI Legacy Survey images with the $r$-band surface brightness limit of $28.5 \rm ~mag/arcsec^2$). These features suggest that tidal interactions may influence the observed kinematics. As a result, the slope of the fitted line in Figure \ref{Fig8}(e) does not provide a reliable measurement of $\Omega_{\rm bar}$ for NGC 7716 (though our slope estimates are still listed in Table \ref{Table3}.) Thus, NGC 7716 is excluded from the subsequent discussion of fast and slow bars, and will instead be discussed separately in Section \ref{chap5.2}.

%and further discussed in Section \ref{chap5.3}.}

%Additionally, the bottom two panels on the right also show a slight redshift, corresponding to two rightmost slits in NGC 7716. 

\subsection{Corotation Radius and Rotation Rate} \label{chap4.3}

\begin{figure}
\centering
\includegraphics[width= 0.95 \linewidth, trim = {10 550 340 50}, clip]{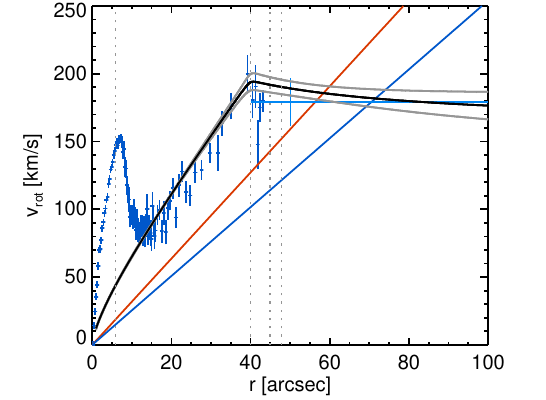}
\caption{Blue-shifted rotation curve of NGC 6951 obtained by fitting pPXF to the long-slit spectrum along the major axis. Uncertainties are estimated using Monte Carlo simulations with 100 iterations. The cyan horizontal line represents the expected flat rotation with a 1 $\sigma$ deviation, while the black solid curve shows the rotation curve modeled with a multi-parameter function after subtracting the nuclear region, which is modeled with a Gaussian. The ranges of uncertainties are indicated by the gray solid curves, obtained using Monte Carlo simulations with 1000 iterations. The red and blue solid lines illustrate the bar rotation fitted with a one- and two-component models for $\Omega_{\rm bar}$. The gray dotted vertical lines indicate the sizes of structures, $R_{\rm ND}$, $R_{\rm \epsilon}$, $R_{\rm Qb}$, and $R_{\rm A2}$, ordered from the center outward.} 
\label{Fig9}
\end{figure}

\begin{figure}
\centering
\includegraphics[width=0.95 \linewidth, trim = {281 550 90 50}, clip]{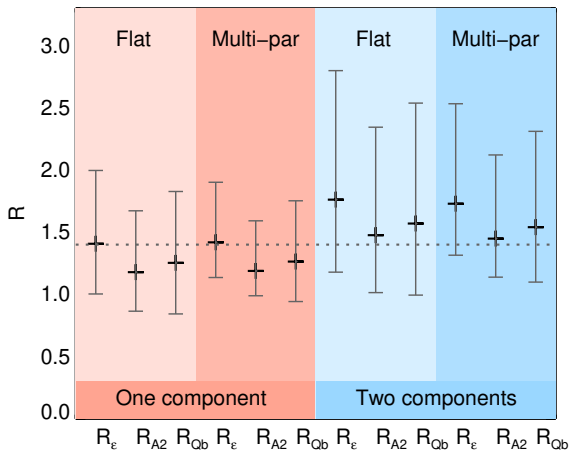}
\caption{$\mathcal{R}$ values for NGC 6951 as functions of the methods used to determine $R_{\rm bar}$, $\Omega_{\rm bar}$, and $v_{\rm rot}$. The $x$-axis indicates the different methods for measuring $R_{\rm bar}$. Reddish and bluish colors represent the one- and two-component models for $\Omega_{\rm bar}$, respectively, while lighter and darker shades indicate flat rotation and multi-parameter fitted rotation curves. Uncertainties are propagated from the errors in $R_{\rm CR}$ and $R_{\rm bar}$.}
\label{Fig10}
\end{figure}

%%---- Table3 ---------------
\begin{deluxetable*}{clrrrr}
 \tablecaption{{Corotation radius and rotation rate in NGC 6951}
 \label{Table4}}
 \tablehead{
 \colhead{$\rm \Omega_{\rm bar}$} &
 \colhead{$v_{\rm rot}$} &
 \colhead{$R_{\rm CR}$} &
 \multicolumn{3}{c}{$\mathcal R$} \\
 \colhead{} &
 \colhead{} &
 \colhead{(arcsec)} &
 \colhead{$R_{\rm CR}/R_{\epsilon}$} &
 \colhead{$R_{\rm CR}/R_{\rm A2}$} &
 \colhead{$R_{\rm CR}/R_{\rm Qb}$} \\
}
\startdata
One component & flat & $56.3^{+23.3}_{-14.8}$ & $1.41^{+0.59}_{-0.40}$ & $1.18^{+0.49}_{-0.31}$ & $1.25^{+0.57}_{-0.41}$ \\
         & multi-par & $56.7^{+19.0}_{-9.3}$ & $1.42^{+0.48}_{-0.28}$ & $1.19^{+0.40}_{-0.20}$ & $1.26^{+0.49}_{-0.32}$ \\
Two components & flat & $70.5^{+41.2}_{-21.9}$ & $1.76^{+1.04}_{-0.58}$ & $1.48^{+0.87}_{-0.46}$ & $1.57^{+0.97}_{-0.58}$ \\
         & multi-par & $69.1^{+31.9}_{-14.5}$ & $1.73^{+0.80}_{-0.41}$ & $1.45^{+0.67}_{-0.31}$ & $1.54^{+0.77}_{-0.44}$ \\ 
\enddata
\end{deluxetable*}
%\vspace{-5mm} % 간격을 5mm 줄임
%----------------------------

Fast and slow bars are commonly distinguished using the rotation ratio, $\mathcal R =R_{\rm CR}/R_{\rm bar}$ \citep{2003Aguerri, 2015Aguerri, 2019Guo, 2019Cuomo, 2020Cuomo, 2020Garma-Oehmichen, 2022Garma-Oehmichen, 2023Geron}, with $R_{\rm CR}$ denoting the corotation radius, the location at which stars rotate at the same angular velocity as $\Omega_{\rm bar}$. A value of $\mathcal R \rm = 1.4$ serves as the threshold, defining fast bars as $1 <\mathcal R \rm \leq 1.4$ and slow bars as $\mathcal R \rm > 1.4$, based on the simulation by \citet{2000Debattista} using halos with different densities. $R_{\rm CR}$ is typically calculated under the assumption of flat rotation \citep{2015Aguerri, 2019Guo, 2019Cuomo, 2024Cuomo}, while some studies have utilized rotation curves fitted with various functions, including arctan and tanh functions \citep{1997Courteau, 2020Garma-Oehmichen, 2021Yoon, 2023Geron, 2025Jeong}. 

Figure \ref{Fig9} represents the rotation velocity of NGC 6951, obtained from pPXF fitting applied to the same spectrum used in Figure \ref{Fig4}. To extend the rotation curve to larger radii, the spectrum was binned to achieve an S/N of 10. The rotation curve on the blue-shifted side provides data at greater distances. As discussed in Section \ref{chap3.2}, NGC 6951 hosts a dynamically decoupled nuclear disk superposed on the main disk. Stars in the nuclear disk rotate faster than stars in the main disk at the same radius \citep{2020Gadotti}. Therefore, the rotation curve of NGC 6951 cannot be effectively modeled using an arctan function. We instead modeled it with two components: a Gaussian fit for the nuclear region and a multi-parameter function \citep{1997Courteau} for the residuals, as shown by the black solid line in Figure \ref{Fig9}. The multi-parameter function is described as follows:
\begin{eqnarray}
v_{\rm rot} = v_{\rm sys}+v_{c}\frac{{1+x}^\beta}{({1+x^\gamma})^{1/\gamma}} 
\end{eqnarray}
where $x=1/R=r_t/r-r_0$. Here, $v_{\rm sys}$ represents the systemic velocity, and $r_0$ is the galaxy center. $v_c$ denotes the asymptotic velocity, while $r_t$ is the transition radius between the rising and flat parts of the rotation curve.
The term $\gamma$ controls the sharpness of the turnover, and $\beta$ accounts for the drop-off or steady rise in the outer part of the rotation curve \citep{1997Courteau}. Uncertainties are estimated using Monte Carlo simulations with 1000 iterations, as indicated by gray solid curve in Figure \ref{Fig9}.

The anticipated flat rotation is depicted by the cyan solid horizontal line, with a value of $179\pm18~\rm km~s^{-1}$, calculated as the average velocity from the maximum near $40\arcsec$ to the endpoint of the obtained curve, with the uncertainty representing the 1$\sigma$ error. Previous studies based on $\rm H\alpha$ observations have reported a similar flat rotation trend in NGC 6951 beyond $50\arcsec$ \citep{1993Marquez, 2002Rozas, 2009Haan}. 

The bar rotation is also shown in Figure \ref{Fig9}, where the red and blue solid lines represent the one-component and two-component models for $\Omega_{\rm bar}$, respectively, as discussed in Section \ref{chap4.2.1}. $R_{\rm CR}$ is determined at the intersection of the bar rotation (red or blue solid line) and the rotation curve (black or cyan solid line). We derived errors of $R_{\rm CR}$ by accounting for uncertainties of both $\Omega_{\rm bar}$ and $v_{\rm rot}$.

The rotation rate $\mathcal R$ is then calculated as the ratio of $R_{\rm CR}$ to $R_{\rm bar}$. Uncertainties are estimated by propagating the uncertainties in both $R_{\rm CR}$ and $R_{\rm bar}$, as summarized in Table \ref{Table4}. Figure \ref{Fig10} visualizes the data from Table \ref{Table4}, showing how $\mathcal{R}$ varies depending on whether a one- or two-component model for $\Omega_{\rm bar}$ is used, as well as on the methods applied to measure $R_{\rm bar}$. The flat rotation and the rotation curve derived using a multi-parameter function yield similar results. When we adopt the criterion of $\mathcal R = 1.4$, NGC 6951 can be classified as a slow bar with the two-component bar model. In the case of the one-component bar model, only the ellipse fitting method for $R_{\rm bar}$ barely falls within the slow bar regime. Further discussions will be provided in Section \ref{chap5.3}.
%The rapidly increasing blue dotted line represents the pattern speed of \textcolor{orange}{a nuclear structure other than the primary bar.}

\section{Discussion}\label{chap5} 

%\subsection{Uncertainties of bar pattern speed measurements} \label{chap6.1}

\subsection{Bar-driven Nuclear Stellar Disk in NGC 6951} \label{chap5.1}

Figure \ref{Fig4} demonstrates the presence of a nuclear stellar disk in each of our target galaxies. In general, the $\sigma$ profiles of galaxies follow an exponential law, peaking at the nucleus, where stars near the bottom of the potential well are expected to have the highest $\sigma$ \citep{2001Emsellem, 2008Comeron, 2017Portaluri}. However, a central decrease in $\sigma$, known as a $\sigma-$drop, is observed in about 50\% of disk galaxies when the spectroscopic data with high spatial resolution are available \citep{2001Emsellem, 2003Marquez, 2006Falcon-Barroso, 2008Comeron, 2014Mendez-Abreu}. The $\sigma$-drop is understood as evidence for the presence of a cold nuclear disk with larger rotational support than the underlying main disk \citep{2003Wozniak, 2006Wozniak, 2017Portaluri, 2020Gadotti}, often associated with nuclear rings, nuclear bars, pseudobulges, or AGN activities \citep{2001Emsellem, 2003Marquez, 2008Comeron, 2017Portaluri}. 

Nuclear disks are believed to form through star formation by dissipative gas funneled into the central regions, primarily driven by bars  \citep{1980Sanders, 1980Simkin, 1992Athanassoula, 2005Ann, 2013Athanassoula, 2014Cole, 2019Seo, 2020Gadotti}. External gas accretion or mergers can also transport gas to the center, leading to nuclear disk formation \citep{2008Mayer, 2011Eliche-Moral, 2013Chapon}, but these mechanisms may leave distinct imprints on the resulting nuclear disk \citep{2011Eliche-Moral, 2020Gadotti}.

%First, nuclear disks formed through dry mergers are distinct from bar-driven nuclear disks, as they exhibit an increasing $\sigma$ and a correlation between $h_3$ and $\sigma$ in their central regions \citep{2011Eliche-Moral, 2020Gadotti}. In contrast, nuclear disks formed through gas-rich mergers show a $\sigma$-drop \citep{2004Morelli}, but \textcolor{orange}{they tend to be smaller, ranging from $60$ to $100~ \rm pc$,} compared to bar-driven nuclear disks \citep{2008Mayer, 2013Chapon, 2014Cole, 2020Gadotti}. Moreover, external processes often lead to counter-rotating nuclear disks \citep{2004Morelli, 2011Eliche-Moral} and are primarily associated with ellipticals, lenticulars, and early-type spirals \citep{1992Rix, 1998Mehlert, 2006McDermid, 2008Krajnovic}, as both counter-rotating gas and the merger process contribute to bulge formation or morphological transformation \citep{2009Scannapieco, 2011Pichon, 2012Sales, 2015Zolotov, 2019Park, 2023Lee}.

The Time Inference with MUSE in Extragalactic Rings (TIMER) project \citep{2019Gadotti} reported that bar-driven nuclear disks exhibit characteristic features including a $\sigma$-drop, an anticorrelation between $h_3$ and $v$, or elevated $h_4$ \citep{2020Gadotti}. They also showed that bar-driven nuclear disks range in size ($R_{\rm kin}$) from $200$ to $1000~\rm pc$, which is strongly related to bar properties, including bar length, strength, and the bar-to-total ratio \citep{2020Gadotti}.

%with $v/\sigma$ values between $1$ and $3$
%These observed features suggest that nuclear stellar disks are kinematically decoupled and rapidly rotating in near-circular orbits.  
%textcolor{orange}{, although} the kinematic axis of the nuclear disk is well aligned with the main disk \citep{2020Gadotti}.

%\begin{figure*}
%\centering
%\includegraphics[width= 0.8 \linewidth, trim = {0 550 120 30}, clip]{Fig10_5.4.3.RC_fitting.SN10.250203.pdf}
%\caption{Rotation curves for NGC 6951 (left) and NGC 7716 (right), derived by fitting pPXF to the binned long-slit spectra while achieving an S/N of 10 along the major axis. The blue-shifted side for NGC 6951 and the red-shifted side for NGC 7716 are used to extend the rotation curves to larger radii. Both curves are well fitted by a simple sum (thick solid line) of two components: a Gaussian profile and a multi-parameter function, with each component represented by thin solid lines. Uncertainties are estimated using Monte Carlo simulations. The sizes of nuclear disks ($R_{\rm kin}$) measured in Figure \ref{Fig4} are indicated by red solid lines, positioned near the peak of the Gaussian profile.}
%\label{Fig10}
%\end{figure*}

In one of our target galaxies, NGC 6951, we find clear evidence for a bar-driven nuclear disk, including a $\sigma$-drop, an anticorrelation between $h_3$ and $v$, and a central increase in $h_4$, as shown in the left column of Figure \ref{Fig4}. The measured values of $R_{\rm kin}$ and $v/\sigma$ are $590~\rm pc$ and $2.8$, respectively, as listed in Table \ref{Table2}. These values fall within the typical range of observed bar-driven nuclear disks \citep{2020Gadotti}. Compared with the bar length and strength in Table \ref{Table2}, the nuclear disk of NGC 6951 follows the expected trends between $R_{\rm kin}$ and bar properties as illustrated in Figure 8 of \citet{2020Gadotti}. Interestingly, this nuclear disk appears as a superposed feature on the rotation curve in Figure \ref{Fig9}, which may result from the superposition of different LOSVDs. This characteristic aligns with the elevated $h_4$ observed in Figure \ref{Fig4}(d). 

%Figure \ref{Fig10} \textcolor{orange}{re}presents the same rotation curve of NGC 6951 (left panel) as in Figure \ref{Fig9} but displays both components used to model it. \textcolor{orange}{This is well described by a simple sum (thick solid line) of a Gaussian profile in the central region and a multi-parameter function representing the underlying main contribution, with each component marked by thin solid lines.} The presence of two components in the central region of NGC 6951 has also been suggested based on the application of the cross-correlation function to the Ca II triplet absorptions \citep{2000Perez}. 
%However, this feature has not been detected in rotation curves derived from $\rm H\alpha$ \citep{1993Marquez, 2002Rozas, 2009Haan, 2000Perez}.

%\textcolor{orange}{Regarding the alignment of the kinematic axes between the nuclear disk and the main disk, although this cannot be directly measured from long-slit observations, the major axis of the nuclear disk in NGC 6951 appears well aligned with that of the main disk within 10 degrees.} Therefore, our analysis consistently supports the presence of a bar-driven nuclear stellar disk in the central region of NGC 6951.

Nuclear disks often contain substructures such as nuclear rings, bars, or spiral arms \citep{2000Ann, 2020Gadotti}. Nuclear rings are closely associated with bar properties, showing that their size is correlated with bar length but an anticorrelation with bar strength \citep{2005Knapen, 2010Comeron, 2012Kim, 2015Li, 2018Sormani, 2020Gadotti}. These relations are similar between nuclear disks and bar properties; this suggests a  fundamental process in common, which is bar-driven \citep{2020Gadotti}. Specifically, nuclear rings ($R_{\rm NR}$) are typically located on the outer edge of nuclear disks ($R_{\rm ND}$) and are considered to be part of them \citep{2014Cole, 2020Gadotti, 2020Bittner}. The left column of Figure \ref{Fig4} shows that $R_{\rm NR}$ (red dotted line) and $R_{\rm ND}$ (red solid line) are consistent in NGC 6951 as well.

%\textcolor{orange}{Lastly, we observe a superposed feature on the rotation curve in Figure \ref{Fig9}, which may be attributed to the superposition of different LOSVDs. This characteristic is consistent with the elevated $h_4$ observed in Figure \ref{Fig4}(d)}. Figure \ref{Fig10} \textcolor{orange}{re}presents the same rotation curve of NGC 6951 (left panel) as in Figure \ref{Fig9} but displays both components used to model it. \textcolor{orange}{This is well described by a simple sum (thick solid line) of a Gaussian profile in the central region and a multi-parameter function representing the underlying main contribution, with each component marked by thin solid lines.} The presence of two components in the central region of NGC 6951 has also been suggested based on the application of the cross-correlation function to the Ca II triplet absorptions \citep{2000Perez}. 
%However, this feature has not been detected in rotation curves derived from $\rm H\alpha$ \citep{1993Marquez, 2002Rozas, 2009Haan, 2000Perez}.

\subsection{Large Nuclear Disk and Nuclear Bar in NGC 7716 without a Primary Bar} \label{chap5.2}

NGC 7716 is also likely to host a nuclear disk with near-circular orbits. As shown in the right column of Figure \ref{Fig4} and in Figure \ref{Fig5}, it exhibits a $\sigma$-drop and an anticorrelation between $h_3$ and $v$. However, the nuclear disk in NGC 7716 appears larger than the typical $R_{\rm ND}$ observed in bar-driven nuclear disks. Its size, $1.2~\rm kpc$, slightly exceeds the typical range of $200$ to $1000~\rm pc$ \citep{2020Gadotti}. Using the values of $R_{\rm bar}$ from Table \ref{Table2}, we find that the ratio $R_{\rm ND}/R_{\rm bar}$ ranges from 26\% to 37\%, depending on the method used to measure $R_{\rm bar}$. This is significantly higher than the typical value of less than 10\% \citep{2020Gadotti}. 

%Although no significant increase in $h_4$ is observed in the central region, its rotation curve is well described by the simple sum of two components, similar to that of NGC 6951, as shown in Figure \ref{Fig10}. Additionally, the major axis of the nuclear disk is aligned with that of the main disk within 5 degrees.

% a superposed structure in its central region,
%However, in NGC 7716, the region of the $\sigma$-drop is very small compared to the size of the nuclear disk (red solid line), as shown in the right column of Figure \ref{Fig4}. 

For the nuclear bar within the nuclear disk in NGC 7716, the length ($R_{\rm NB}$) is measured to be $810~\rm pc$ based on visual inspection of the unsharp mask, as shown in Figure \ref{Fig3}. While $R_{\rm NB}$ is generally positively correlated with the stellar mass of the host galaxy \citep{2024Erwin}, the nuclear bar in NGC 7716 is significantly longer than expected from the nuclear bar–stellar mass relation. Instead, it falls within the lower regime of the primary bar–stellar mass relation, suggesting that its size is more comparable to a short primary bar rather than a typical nuclear bar \citep[see Figure 9 in][]{2024Erwin}, assuming a stellar mass of $log~\mathrm{M_{\ast}} = 10.17 \pm 0.08 ~\mathrm{M_{\odot}}$ \citep{2018Aquino-Ortiz}.

On the other hand, \citet{2020deLorenzoCarceres} reported a bimodal distribution of $R_{\rm NB}/R_{\rm bar}$, suggesting distinct formation or evolutionary pathways. For instance, based on the TNG50 simulations, \citet{2024Semczuk} proposed that nuclear bars may form first, either in isolation via instabilities or through galaxy interactions. This mechanism could explain how nuclear bars grow longer relative to the primary bar, as observed in \citet{2020deLorenzoCarceres}. Notably, NGC 7716 hosts a large nuclear bar, with $R_{\rm NB}/R_{\rm bar}\sim$ 18\%$-$26\%, depending on the method used to measure $R_{\rm bar}$. These values could increase when using the decomposition method employed by \citet{2020deLorenzoCarceres}. 

When it comes to the primary bar of NGC 7716, it appears as an oval structure surrounded by an inner ring (Figure \ref{Fig1}(b)) and has been visually classified as a weak bar \citep{1991deVaucouleurs}. This classification is supported by the ellipse fitting method (Figure \ref{Fig7}(c)). However, Fourier analysis and the force ratio map (Figures \ref{Fig7}(d) and (e)) make it challenging to classify NGC 7716 even as a weak bar. In particular, the bar strength is notably low, with $Q_{\rm b} = 0.07$, which is well below the typical threshold of $Q_{\rm b} = 0.15$. Additionally, the minimal contribution of non-axisymmetric components to the luminosity within each long-slit spectrum further supports the weakness of the bar structure (Figure \ref{Fig8}(d)). 

Therefore, the growth of a large nuclear bar in NGC 7716 may be facilitated by the absence or weakness of the primary bar, similar to the longer nuclear bars observed in the TNG50 simulations \citep{2024Semczuk}. Considering the tidal stream revealed in the deep image \citep{2018Bakos}, NGC 7716 may have undergone a more complex evolutionary history.

%\subsection{Detection of a Slow Bar in NGC 6951 with a Nuclear Ring} \label{chap5.2}
\subsection{Two Pattern Speeds in NGC 6951} \label{chap5.3}

The possibility of two pattern speeds in NGC 6951, as discussed in Section \ref{chap4.2.1}, is also an intriguing aspect of this study. Previous N-body simulations have shown that galaxies can host independent structures, such as bars, spirals, and nuclear bars, each rotating with a different pattern speed \citep{1988Sellwood, 1999Rautiainen, 2006Minchev, 2012Minchev, 2015Baba, 2024Zouloumi}. This finding is supported by observations detecting multiple $R_{\rm CR}$ \citep{1997Puerari, 1998Aguerri} or multiple pattern speeds \citep{2009Meidt}. Slowly rotating spiral arms compared to a bar have been observed in the Milky Way \citep{2005Dias, 2011Gerhard}, and in other spirals \citep{2016Speights}.

In the TW method, a different pattern speed appears as a slope change in the straight-line fitting to the TW integrals \citep{2003Corsini, 2021Cuomo}. A rapidly rotating nuclear bar was first observed in NGC 2950 using the TW method \citep{2003Corsini}. In Figure \ref{Fig8}(b), we observe a slope change similar to that seen in NGC 2950, suggesting the presence of two pattern speeds in the central region of NGC 6951. Considering the offset between slits ($11.\!\arcsec45$), we expect the presence of a bar-like structure around $r \sim 10\arcsec$. We also found several pieces of evidence supporting for this structure. Firstly, the $\epsilon$ and PA profile show a bar-like structure at a comparable radius in both HST/PC and PS1 photometry (Figures \ref{Fig3}(c-d) and \ref{Fig6}(c)); Fourier analysis further supports this feature (Figure \ref{Fig6}(d)). Secondly, the rapidly rotating pattern speed converges when the slit length approaches $r \sim 10\arcsec$, similar to the convergence of $\Omega_{\rm bar}$ near $R_{\rm bar}$. Nevertheless, neither the PC image nor the corresponding unsharp masked image (Figures \ref{Fig2}(b) and (f)) provides clear evidence of a bar-like structure at this radius. 

Similarly, several previous studies anticipated the presence of a nuclear bar in NGC 6951 but were unable to detect it \citep{1996Friedli, 2000Perez, 2002Rozas, 2009Haan}. \citet{2000Perez} suggested the possibility of a gas-driven structure, such as a nuclear bar, based on the relatively high molecular-to-total mass fraction within $r \lesssim 6\arcsec$, which is consistent with the size of the kinematic nuclear disk, $R_{\rm kin} = 5.\!\arcsec9$, measured in Figure \ref{Fig4}(e). \citet{2002Rozas} and \citet{2009Haan} inferred it from the detection of strong non-circular motion near the nucleus. Therefore, the bar signature at $r \sim 10\arcsec$ and the two pattern speeds observed in NGC 6951 in this study are consistent with the properties reported in previous studies \citep{2000Perez, 2002Rozas}. Instead of a bar-like structure at $r \sim 10\arcsec$, previous studies have reported nuclear spirals and a nuclear oval within the nuclear ring \citep{2000Perez, 2005Garcia, 2007Storchi, 2011vanderLaan}, which are likely smaller in size given the slit positions. Although we do not observe a distinct feature in the images, our results suggest that an oval structure of $r \sim 10\arcsec$ could be responsible for inducing the rapidly rotating pattern speed.

\subsection{Slow, Long, and Strong Bar in NGC 6951} \label{chap5.4}

When we adopt the two pattern speeds of NGC 6951, the galaxy is classified as a slow bar ($1.45 \leq \mathcal{R} \leq 1.76$ in Table \ref{Table4}). Moreover, all three methods used to measure bar strength, ellipse fitting, Fourier analysis, and the force ratio map, consistently indicate that NGC 6951 hosts a strong bar ($\epsilon_{\rm max} = 0.55$, $A_2 = 0.42$, $Q_{\rm b} = 0.36$), despite its visual classification as a weak bar, as discussed in Section \ref{chap4.1}.

%Typically, the bar lengths measured using Fourier analysis and the force ratio map are comparable and tend to be approximately 20$\%$ shorter than those measured by the ellipse fitting \citep{2022Lee}. Since the criterion of $\mathcal{R} = 1.4$ was established through Fourier analysis in the simulations of \citet{2000Debattista}, NGC 6951 meets the classification of a slow bar. This implies that when bar lengths are measured using ellipse fitting methods, some slow bars may be classified as fast bars according to the criterion of $\mathcal{R}~(= R_{\rm CR}/R_{\rm bar}) = 1.4$ \citep{2021Cuomo, 2022Lee}.

 On the other hand, assessing whether a bar is long is somewhat complex. For instance, while visual inspection often relies on relative bar length to distinguish between SBs and SABs \citep{2015Ann, 2015Buta, 2019Lee}, the relative bar lengths measured using the three methods ($R_{\epsilon}/R_{25}$, $R_{A2}/R_{R25}$, and $R_{Qb}/R_{25}$) do not show significant differences in their distributions \citep{2020Lee}. Moreover, bar length depends on morphological type or bulge-to-total, with early-type spirals generally hosting longer bars in both absolute and relative terms \citep{1985Elmegreen, 1995Martin, 2007Menendez-Delmestre, 2009Aguerri, 2011Gadotti, 2014Kim}. More specifically, the relative bar length exhibits a bimodal distribution, increasing toward both ends of the Hubble sequence \citep{2007Laurikainen, 2016Diaz, 2017Font, 2019Font, 2020Lee}. Accordingly, when we compare the relative bar length of NGC 6951 to the distribution of relative bar lengths across the Hubble sequence \citep[Figure 14 in][]{2020Lee}, it is classified as a long bar in an Sbc galaxy \citep{1991deVaucouleurs}. The relative bar lengths of NGC 6951, measured using three different methods, are higher than the mean values for SBbc galaxies. In particular, $R_{A2}/R_{25}$ and $R_{Qb}/R_{25}$ exceed the mean by nearly a factor of two.

 As a result, based on all analyses presented in this study, we classify NGC 6951 as a galaxy with a slow, long, and strong bar. Additionally, NGC 6951 is a deeply isolated galaxy, with no nearby galaxies within 1 Mpc in projected distance, implying no significant gravitational influence from other galaxies for at least past $10^9$ years \citep{1993Marquez}. Therefore, we suggest that NGC 6951, with a nuclear ring, could be an ideal candidate for a slow, long, and strong bar evolved through interactions with a dark matter halo, as predicted in simulation studies \citep{1980Sellwood, 2000Debattista, 2002Athanassoula, 2003Athanassoula, 2014Athanassoula, 2019Seo, 2023Jang, 2024Jang}. This slow, long, and strong bar may significantly contribute to the formation of the nuclear structure, including its nuclear disk and nuclear ring.

\section{Conclusions}\label{chap6}
In this work, we observed two barred galaxies with nuclear structures in order to search for examples of slow bars evolved through interactions with their dark matter halos. Using high-resolution spectroscopy from Gemini/GMOS long-slit observations and high-resolution photometry from HST/PC and PS1 archival data, we investigated the properties of their nuclear structures and primary bars, including bar pattern speed ($\Omega_{\rm bar}$), length ($R_{\epsilon}$, $R_{A2}$, and $R_{Qb}$), strength ($\epsilon_{\rm max}$, $A_2$, and $Q_b$), corotation radius ($R_{\rm CR}$), and rotation rate ($\mathcal{R}$).

Our main results are as follows.
\begin{enumerate}
\item High-resolution spectroscopy reveals kinematically decoupled nuclear disks in both target galaxies, while high-resolution photometry identifies a nuclear ring in NGC 6951 and a nuclear bar in NGC 7716. These nuclear structures are embedded within their respective nuclear disks.
\item Our findings suggest that NGC 6951 hosts a slow, long, and strong bar, which likely evolved through interactions with the dark matter halo. Both the nuclear disk and the nuclear ring in NGC 6951 appear to have formed through a common bar-driven mechanism. Additionally, we identify a rapidly rotating pattern speed within the primary bar, which may be driven by an embedded oval structure, although this feature is not clearly visual in the imaging data.
\item For NGC 7716, we observe unexpected results. The primary bar is too weak to be classified as barred, and we were unable to determine its bar pattern speed due to redshifted velocities observed in certain slits, which may be affected by external sources. Nevertheless, NGC 7716 hosts a large nuclear disk and nuclear bar. We suggest that the formation of its long nuclear bar may follow a different evolutionary pathway, potentially influenced by tidal interactions or by the weakness of the primary bar.
\end{enumerate}

As a result, our findings suggest that slow bars may be more observed in galaxies with nuclear structures, which may have developed through interactions with dark matter halos. Furthermore, galaxy interactions may play a frequent and significant role in galaxy evolution, potentially affecting even the formation of nuclear structures.

\section*{Acknowledgement}
We sincerely thank the anonymous referee for the detailed and insightful comments, which greatly helped improve the paper. This research was supported by Basic Science Research Program through the
National Research Foundation of Korea (NRF) funded by the Ministry of Education (No. RS-2023-00249435) and by the Korean government (MSIT; No. 2022R1A4A3031306). HSH acknowledges the support of Samsung Electronic Co., Ltd. (Project Number IO220811-01945-01), the National Research Foundation of Korea (NRF) grant funded by the Korea government (MSIT; NRF-2021R1A2C1094577), and Hyunsong Educational \& Cultural Foundation. V.C. acknowledges support from the Agencia Nacional de Investigación y Desarrollo (Chile) through the FONDECYT Iniciación Grant (No. 11250723).
M.G.P. acknowledges the support from the Basic Science Research Program through the National Research Foundation of Korea (NRF) funded by the Ministry of Education (No. RS-2019-NR045193 and RS-2018-NR031074). T.K. acknowledges support from the Basic Science Research Program through the National Research Foundation of Korea (NRF) funded by the Ministry of Education (No. RS-2023-00240212). NH, H-J.K., and JYS acknowledge the support by by the Korea Astronomy and Space Science Institute grant funded by the Korea government (MSIT; No. 2025-1-860-02, International Optical Observatory Project). The work of W.-T.K. was supported by the National Research Foundation of Korea (NRF) grant funded by the Korea government (MSIT; No. RS-2025-00517264). J.H.L. acknowledges support from the Basic Science Research Program through the National Research Foundation of Korea (NRF) funded by the Ministry of Education (No. RS-2024-00452816).

This work was supported by K-GMT Science Program (PID: GN-2023B-Q-303) of Korea Astronomy and Space Science Institute (KASI). Based on observations obtained at the international Gemini Observatory, a program of NSF NOIRLab, which is managed by the Association of Universities for Research in Astronomy (AURA) under a cooperative agreement with the U.S. National Science Foundation on behalf of the Gemini Observatory partnership: the U.S. National Science Foundation (United States), National Research Council (Canada), Agencia Nacional de Investigaci\'{o}n y Desarrollo (Chile), Ministerio de Ciencia, Tecnolog\'{i}a e Innovaci\'{o}n (Argentina), Minist\'{e}rio da Ci\^{e}ncia, Tecnologia, Inova\c{c}\~{o}es e Comunica\c{c}\~{o}es (Brazil), and Korea Astronomy and Space Science Institute (Republic of Korea).


\vspace{5mm}
\facilities{Gemini, PS1, SDSS, HST}


\software{Dragons \citep{2019Labrie}, pPXF \citep{2004Cappellari}, BarIstA \citep{2019Lee, 2020Lee}}

%\appendix

\bibliography{24_nuclear_structures}{}
\bibliographystyle{aasjournal}






\end{document}